%% file: mobihoc_2019_arxiv.tex
\documentclass[journal]{IEEEtran}
\usepackage{hyperref}
\usepackage{amsmath,amsthm}
\usepackage{graphics, subfigure}
\usepackage{tikz}
\usepackage{mathtools}
\usetikzlibrary{shapes,arrows}
\usepackage{algorithm, algorithmic}
\usepackage[outdir=./]{epstopdf}
\usepackage{color}
\usepackage{multirow}
\usetikzlibrary{automata,positioning}
\usepackage{wrapfig}
\usepackage{rotating}

\usepackage{enumitem}

\newtheorem{assumption}{Assumption}
\newtheorem{assum}{Assumption}
\newtheorem{theorem}{Theorem}
\newtheorem{lemma}{Lemma}

\newtheorem{definition}{Definition}

\newtheorem{cor}{Corollary}
\newtheorem{corollary}{Corollary}
\newtheorem{lem}{Lemma}
\title{Competition with Three-Tier Spectrum Access and Spectrum Monitoring}

\author{Arnob Ghosh\thanks{Arnob Ghosh is with the Industrial Engineering Department of Purdue University, His e-mail id is ghosh39@purdue.edu}, Randall Berry\thanks{Randall Berry is with the Electrical Engineering and Computer Science Department of Northwestern University. His e-mail id is rberry@eecs.northwestern.edu}}
\begin{document}
\maketitle

\begin{abstract}
\small The  Citizens Broadband Radio Service (CBRS) recently adopted in the  U.S.
enables two tiers of commercial users to share spectrum with a third tier of incumbent users. This sharing can be further assisted by Environmental Sensing Capability operators (ESCs), that monitor the spectrum occupancy to determine when use of the spectrum will not harm incumbents. Two key aspects of this framework that impact how firms may compete are the differences in information provided by different ESCs and the different tiers in which a firm may access the spectrum. We develop a game theoretic model that captures both of these features and analyze it to gain insight into their impact. Specifically, we consider a priority access (PA) tier  firm has access to the both licensed band and unlicensed band, and a general authorized access (GAA)  tier firm has access only to the unlicensed band. The PA tier and GAA tier firms compete for users. 
Our analysis reveals that the amount of unlicensed and licensed bandwidth in the CBRS must be chosen judiciously in order to maximize the social welfare. We also show that a limited amount of unlicensed access of PA tier firm is beneficial to the user's surplus as well as to the social welfare. 
%Our analysis reveals several counter intuitive results.
\end{abstract}

\input{intro}
\input{system}

In the following, we, first, describe the scenario where only one of the SAs obtain information from an ESC. Subsequently, we consider the scenario where both the SAs obtain information from the same ESC. Finally, we consider scenarios where the SAs obtain information from different ESCs.  In each case, we characterize the market equilibrium in the second and third stage via backward induction.   
\input{monopoly}
\input{same_esc}
\input{diff_esc}
\input{first_stage}

\input{numerical}

\section{Conclusion}
We studied a model motivated by the emerging CBRS standard which shows that differences in spectrum access tiers and information about spectrum availability  can have a large impact on the markets that may emerge.  There are many ways this could be extended including modeling more firms or the strategic pricing of the ESCs. Consideration of a dynamic market where the SAs can obtain information from the ESCs in a spot market is also left for the future.

\newpage
\clearpage
\input{proof}
\end{document}

%% file: intro.tex
\section{Introduction}
Recently, the U.S. FCC has finalized plans for the Citizens Broadband Radio Service (CBRS)  \cite{FCC-CBRS}. These plans enable commercial users to use the 3.5GHz band (3550-3700 MHz) when incumbent users (e.g., federal users and  fixed satellite users) are absent.   Accessing this band in a given location is to be controlled by one or more {\it Spectrum Access Systems} (SASs), which are geographic databases that contain information about the locations and spectrum utilization of users of this spectrum.  It is envisioned that in many areas, multiple companies will operate approved SASs.\footnote{In the first round of applications, six different companies requested FCC approval as SAS operators \cite{FCC-SAS-16}.}
Companies wishing to offer service in that band must then register with one SAS. Additionally, each SAS can utilize an {\it Environmental Sensing Capability operator} (ESC).  An ESC will utilize a network of sensors to detect the presence (or absence) of federal incumbent users, enabling firms to better utilize the spectrum than would be possible under more conservative exclusion zones. 

It is also envisioned that multiple ESCs may be present in the market.\footnote{For example, as of February, 2018, four different entities have been conditionally approved as ESC operators\cite{FCC-ESC-18}.} If these ESCs offer different qualities of spectrum measurements, this in turn could impact the downstream competition of wireless {\it spectrum access firms} (SAs) who are using the spectrum to serve their customers.  A SA contracting with a ``better" ESC will be able to utilize the spectrum more often and thus offer a higher quality of service, giving it a competitive advantage. However, using such an ESC might also entail a higher cost.

 In addition to the information provided by an ESC, another key aspect of CBRS is that there are two different tiers of commercial access allowed: a {\it Priority Access} (PA) tier and a {\it General Authorized Access} (GAA) tier.  The PA tier provides a form of licensed access in that a SA with a {\it Priority Access License} (PAL) is given the exclusive right to use a portion of the CBRS spectrum in a given location when the incumbent users are not present. The GAA tier allows for a type of unlicensed access by SAs: any SA may utilize spectrum that is not needed by an incumbent or PA user. The guidelines for this band also limit the portion of spectrum that can be allocated a PAL, so that when incumbents are not present, a portion of the band will be available for GAA use. 
Thus in a given location there could be SAs that differ in both the tier in which they access spectrum and the ESC from which they acquire information. This raises many interesting questions. 
Can different quality ESCs co-exist in a given market and if so how do difference in their quality impact the resulting market?  How does the tier at which a SA accesses spectrum impact the ESC  quality it is willing to accept?  How does this impact the downstream competition among SAs? How much of the bandwidth should be reserved for the PA tier?
In this paper, we study  a stylized model of the CBRS ecosystem to gain insight into these questions.

We consider a market in which SAs seek to utilize a given band of spectrum in a given area which is shared in a manner similar to that in the CBRS system.  To utilize the spectrum each SA needs to register with a SAS operators serving the given area so as to obtain information regarding spectrum availability at a given cost. We assume that each SAS operator utilizes a different ESC to obtain the spectrum measurement data used in its database. For simplicity in the following we refer to the combined ESC/SAS operator as simply an ESC 
operator.\footnote{The combined  ESC/SAS operator could be a single firm that both maintains a SAS and a ESC network or two separate firms that have a contractual agreement.  In this paper we do not address the details of such agreements.} We differentiate the ESCs by both the price they charge a SA and the probability that they indicate that the spectrum is available.  Given their choice of ESC, the SAs in turn compete to serve end-users in a given area whenever the ESC data tells them the spectrum is available.  We capture the differences in the tier in which an SA might operate by assuming that the given band is divided into two portions: a {\it licensed band} modeling the PA tier and an {\it unlicensed band}  modeling the  GAA tier.  To keep our analysis tractable, we focus on a duopoly scenario in which there are only two SAs; however, we note at several places where our analysis could be extended to market with more participants. 
Further, to highlight the impact of different spectrum access tiers, we assume that only one of the SAs has a PAL enabling it to use the licensed portion of this band (in addition to the unlicensed portion), while the other SA can only utilize the unlicensed portion. 

We formulate a multi-stage game in which the SAs first decide on the ESC to contract with and then compete for customers. Our model for the competition among the SA's for customers is based on the literature for price competition with congestable resources, e.g., \cite{EngelFG99,AH03,AO07,HTW05}. In these models firms compete for customers by announcing prices; the customers in turn select firms based on a {\it delivered price} which is the sum of the announced price and a congestion cost that depends on the number of users using a firm's resources. This type of model has been widely used to study competition among wireless service providers including the work in 
\cite{nguyen2011impact,maille2012competition,kavurmacioglu2012competition,zhang2013competition,
nguyen2014cost,liu2014competition,infocom_18}. Our approach is closest to that in  \cite{infocom_18} which considers a scenario in which a service provider can serve users using a combination of licensed and unlicensed spectrum. Similarly, we consider a model in which the SA with a PAL is allowed to use both the licensed and unlicensed portion of the band when it is available. Similar to \cite{infocom_18}, we model this by assuming that this SA's customers are served on the unlicensed band with probability $\alpha$ and on the licensed band with probability $1-\alpha$. An important distinction in our work from all of the above, is that we assume that the given band of spectrum is not always available and that this availability also depends on the information acquired by the ESC. For this we build on our preliminary  work in \cite{icc_18} that also considers a setting where different ESCs sell information to the SAs. However, \cite{icc_18} assumes that all of the spectrum is unlicensed and thus does not address the different access tiers as in our model.  We find that adding these tiers to the model significantly complicates the analysis and leads to very different conclusions.

We initially characterize the pricing equilibrium under different possible ESC selections and subsequently examine the equilibrium ESC selection.  The first ESC selection we consider 
 is when only one SA obtains information from a ESC, while the other SA stays out of the market (Section~\ref{sec:mon}). In this case the one ESC  in the market acts as a monopolist when pricing its services and thus can extract all of the user surplus. We next turn to the case where both SAs obtain information from the same ESC (Section~\ref{sec:same_esc}). In \cite{icc_18}, we showed that when all the spectrum is unlicensed, this type of ESC selection always leads to zero profits for the SAs.  However, in our setting, we show that even if the licensed bandwidth is very small, there exists an $\alpha<1$ that will ensure that the payoff of both SAs are not zero under this ESC selection. Our analysis also reveals that there is a threshold on $\alpha$ beyond which, the SA that only has access to the unlicensed spectrum  (say SA $2$) achieves a negative profit (and so would not want to enter this market). The threshold  decreases as the ratio of the licensed band and unlicensed band increases, i.e., if a larger proportion of the band is allowed to be licensed, then $\alpha$ must be more constrained for SA $2$ to achieve a  positive profit. This suggests that if the goal is ensure greater competition in a market with a single ESC, then perhaps in addition to limiting the proportion of spectrum that is licensed, one should also limit the amount of traffic a license holder can place on the unlicensed band. 

Subsequently, we characterize the equilibrium strategy when the SAs obtain information from different ESCs (Section~\ref{sec:diff_esc}). We show that when the SA with licensed access (say, SA $1$) obtains information from the ``better" ESC, it can drive the profit of SA $2$ to zero when the user valuation for service is large enough. Nevertheless, in this setting we show that when the user valuation is not too high and $\alpha$ is high enough, SA $2$ can attain a strictly higher payoff compared to the scenario where the SAs obtain information from the same ESC.  On the other hand, when SA $2$ 
obtains the better quality of information, we show that if  $\alpha$ is high, the payoffs of both the SAs are always positive. This shows that if the ESCs have different quality, then limiting $\alpha$ may not be needed to promote competition. 

We, finally, characterize the equilibrium  choice of ESCs by the SAs (Section~\ref{sec:first_stage}). Our analysis reveals that if $\alpha=0$ ($\alpha=1$, resp.), the SAs will never obtain information from different ESCs (same ESC, resp,). We show that, surprisingly,  when  $\alpha$ is above a threshold, and the ratio of unlicensed bandwidth to licensed bandwidth decreases to zero, the SAs obtain information from different ESCs (Theorem~\ref{thm:lowu}). When $\alpha$ is below the threshold, a monopoly arises. However, if the ratio of unlicensed bandwidth to licensed bandwidth increases to infinity, the SAs obtain information from the same ESC. Note that if all the bandwidth was unlicensed, as in \cite{icc_18},
there is  {\em no} equilibrium where both the SAs will select the same ESC. In contrast, our analysis reveals that if the licensed band is non-zero, and the unlicensed band is sufficiently large, in the equilibrium the SAs will always select the same ESC.% If the unlicensed bandwidth is too small, and $\alpha$ is small, there may be a monopoly scenario where only the SA which has licensed access will obtain information from a ESC and serve the users. The other SA will not be present in the market. 

We, empirically, characterize the payoffs of the SAs, the user's surplus, and the social welfare as a function of the ratio between the licensed bandwidth and the unlicensed bandwidth, and $\alpha$ (Section~\ref{sec:numerical}). Our analysis shows that user's surplus and social welfare are neither maximized when the licensed bandwidth is too large nor when the licensed bandwidth is too small compared to the unlicensed bandwidth. Thus, a regulator may need to select the proportion of  bandwidth for licensed access  judiciously.  Note that the $\alpha=0$ scenario is equivalent to the scenario where each SA has exclusive access. Thus, the congestion should be minimum when $\alpha=0$. However, interestingly our analysis reveals that $\alpha=0$ may not maximize the social welfare, or the users' surplus. {\em Thus, a limited utilization of the unlicensed bandwidth by PA tier SA increases the social welfare as well as the user's surplus.}

%We focus on a single geographic area and assume both SAs only use a single shared band of spectrum (i.e., neither SA has access to other bands of spectrum).  We focus on the case where the ESCs have different information regarding the presence of the incumbent because of different sensing capabilities. The SAs can obtain information from at most one of the ESCs. If a SA does not obtain information from any of the ESCs, we assume that it can not offer service to the customers (i.e., this could model a situation where the given location is within an exclusion zone).  We then analyze a multi-stage game in which the SAs first decide on contracting with a ESC. Given these decisions, the SAs then compete for users.

%% file: system.tex
\section{System Model}\label{sec:system}
\begin{figure}
\includegraphics[trim=1in 4in 2in 2in,width=0.25\textwidth]{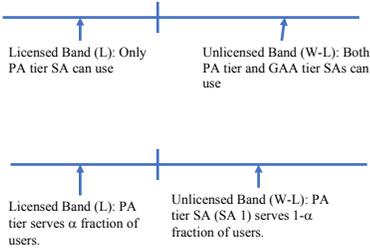}
\vspace{-0.3in}
\caption{The PA tier SA can use the licensed as well as the unlicensed band. The GAA tier SA is only restricted to the unlicensed band.}
\label{fig:pa_gaa}
\vspace{-0.3in}
\end{figure}

We consider a scenario in which two SAs denoted by SA $1$ and SA $2$ seeks to serve users at a given location using a given band of spectrum with bandwidth $W$. This band of spectrum operates under a three-tier spectrum access model in which a bandwidth of $L$ is licensed (i.e., at the PA tier) and the remaining $W-L$ is unlicensed (i.e., at the GAA tier) (Figure~\ref{fig:pa_gaa}.   In order to use the spectrum, each SA must acquire spectrum measurements from one of two ESCs and can only use the spectrum when the ESC indicates that it is available (i.e., it is not used by an incumbent).   We next discuss the market participants in more detail.

\subsection{SAs}

We assume that only SA $1$ has access to the licensed portion of the spectrum (bandwidth $L$), while both SA's can use the unlicensed portion (bandwidth $W-L$).  Hence, SA $1$ can serve users using both the licensed and the unlicensed spectrum (Figure~\ref{fig:pa_gaa}). We model this by assuming that SA $1$ assigns users to the licensed band with probability $1-\alpha$, and the unlicensed band with probability $\alpha$ (Figure~\ref{fig:pa_gaa}). We assume that this decision is made independently for each user in any given time instant, so that over a service period (say one month), each user of SA $1$ will spend approximately $\alpha$ of the time using the unlicensed band and the remaining time using the licensed one. Throughout the analysis, we consider that $\alpha$ is exogenously determined. \begin{color}{black}This could be specified by a regulator or determined by some underlying technology. In other cases, it might be determined by SA $1$ strategically. If $\alpha$ is set by the regulator, the regulator may want to maximize the social welfare. On the other hand, if the SA itself decides $\alpha$ it may want to maximize its own revenue. In Section~\ref{sec:numerical}, we empirically show the values of $\alpha$ which maximize the social welfare and/or the SA's profit. \end{color}

\begin{color}{black}Each SA $i,$ selects a price $p_i$ it will charge users and also decides the ESC $j\in \{A,B\}$ from which it will access information.  As is the case in the wireless market today,  we view the price $p_i$ as representing the amount users pay for receiving long-term service from SA $i$ (e.g.,the  monthly service price). As such these prices represent the service from an SA averaged over this service period. Here we view these as flat-rate prices, and assume that each SA only offers a single service plan (which is reasonable as in our model the user population is homogeneous).
 \end{color}

Let $\lambda_i$ be the number of users of SA $i$. SA $i$ then obtains a revenue of $p_i\lambda_i$ and has to pay a price $\tilde{p}_j$ to the ESC it selects, giving it a total payoff of 
\begin{align}\label{eq:profit}
p_i\lambda_i-\tilde{p}_j.
\end{align}
 If a SA does not obtain information from any ESC, it can not enter the market and so its payoff is zero. 

\subsection{ESCs }
Each ESC $j$ obtains a payment of $\tilde{p}_j$ when a  SA  obtains information from it. If no  SA obtains information from an ESC, the ESC's payoff is zero. In this paper, we focus on the competition between the SAs and so do not consider the ESC as a strategic player. Specifically, we assume that $\tilde{p}_j$ is an exogenous parameter to the model, and it is not a decision variable. 

We assume that the incumbent users utilize the entire band $W$ when present. The ESC in turn provides estimates of the band's availability.  
Each ESC estimates that the band is available with probability $q_j$, $j\in \{A,B\}$ if the incumbent is absent.  We assume that  ESC $B$'s estimation is of inferior quality compared to ESC $A$. Specifically, if ESC $B$ estimates that the channel is available, ESC $A$ also does the same; however, the converse is not true, so that $q_A>q_B$.\footnote{This means that ESC $B$'s estimate is a degraded version of ESC $A$'s estimate.  Alternatively, one can model the ESCs as making independent error, with the respective probabilities. This results in a more cumbersome model and so we focus on the former case here.}  Each ESC has a negligible probability of miss-detection, i.e., if the incumbent is present, the ESC will never announce that the spectrum is available.

\subsection{User's Subscription Model}\label{sec:user}
We consider a mass $\Lambda$ of non-atomic users, so that  $\lambda_1+ \lambda_2 \leq \Lambda$.   The users are assumed to be homogeneous so that each user obtains a value $v$ for getting service from either SA over the service period. However, as in 
\cite{maille2012competition,nguyen2014cost,nguyen2011impact} users also incur a 
{\it congestion cost} when using this service. The congestion cost models the degradation in service due to congestion of network resources.
Here, we model  the congestion cost for using a band of spectrum with bandwidth $B$ by $x/B$ where $x$ is the total mass of users using that band. More generally, the congestion cost could be given by $g(x/B)$, where $g(\cdot)$ is an increasing, convex function. Here, we assume this function is linear to simplify the analysis, similar to \cite{nguyen2011impact}. The dependence on $B$ models the fact that a larger band of spectrum is able to support more users. 

Over a given service period, the congestion cost will vary depending the information available to the SAs and the tier of spectrum a user is assigned to.   We assume that over this period, SA's are sensitive to the expected congestion cost, which is taken by averaging across these parameters. Again, as we are viewing the pricing and service choices of users as being over a long time-horizon, this is a reasonable assumption.   Recall that a user also has to pay the price $p_i$ if it subscribes to SA $i$. The pay-off of a user receiving service from SA $i$ is then given by
\begin{equation}\label{eq:net_ben}
E(v - p_i -x/B),
\end{equation}
where the $E$ indicates that this is evaluated by taking an expectation over the possible realizations of SA information and the spectrum tier assignment. Users can also choose not to purchase service from either SA, in which case their pay-off is zero.

We also define a  user's {\it ex post payoff}  to be the value in (\ref{eq:net_ben} conditioned on a particular realization of the information available to the SAs and the assignment of a user to a tier of spectrum. For example, when a user receives service from SA $i$ and cannot use the spectrum (due the ESC indicating it is not available), its ex post payoff is $-p_i$. In the following, we will discuss the ex post payoff for the other possible realizations. A user's pay-off in (\ref{eq:net_ben}) can then be viewed as the average of the  possible ex-post pay-offs seen by that user (where the probability of these different realizations will depend on the ESC selection choice of the SAs).   

We, first, discuss the ex post payoff attained by the users of SA $1$. If a user of SA $1$  is served using the licensed spectrum its ex post payoff is
\begin{equation}\label{eq:cong_license}
v-p_1-(1-\alpha)\dfrac{\lambda_1}{L}.
\end{equation}
If a user of SA $1$ is served using the unlicensed spectrum, its congestion cost depends on two components. The user will face congestion from the other users of SA $1$ which are being served using the unlicensed spectrum,  and  it will also face congestion from the users  of SA $2$ if SA $2$ also has information that the band is available. In this case, the users of SA $1$  
achieve an ex post pay-off of 
\begin{equation}\label{eq:same_cong}
v-p_1-\dfrac{\alpha\lambda_1+\lambda_2}{(W-L)}.
\end{equation}
If SA $2$ does not know the band is available, the payoff of a subscriber of SA $1$ served in the unlicensed band is\footnote{Note that $\lambda_2$ is absent from the payoff term in (\ref{eq:same_cong}) since SA $2$ thinks the band is unavailable.}
\begin{equation}\label{eq:cong_diff}
v-p_1 - \dfrac{\alpha\lambda_1}{W-L}.
\end{equation}

We next characterize the ex post payoff attained by the users of SA $2$. SA $2$'s users face congestion from the users of SA $1$ that  are served using the unlicensed spectrum if SA $1$ also knows that the spectrum is available, giving a pay-off of
\begin{equation}
v-p_2-\dfrac{\alpha\lambda_1+\lambda_2}{W-L}.
\end{equation}
Note that the congestion cost is similar to (\ref{eq:same_cong}) since in this case both the SAs know that the  incumbent is absent. 
If the SA $1$ does not know that the spectrum is available, SA $2$'s users  only face congestion from themselves, leading to a pay-off of
\begin{equation}\label{eq:cong_diff2}
v-p_2-\dfrac{\lambda_2}{W-L}. 
\end{equation}

%where $x$ is the number of users using the unlicensed spectrum. 

\subsection{Multi-Stage Market Equilibrium}

\begin{figure}
\includegraphics[trim=1in 4in 4in 1in,width=0.25\textwidth]{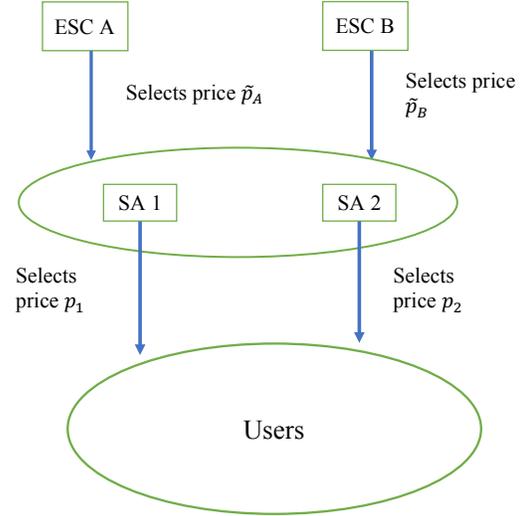}
\vspace{-0.3in}
\caption{In the first stage, the ESCs select prices to the SAs. After SAs choose to obtain information from one of the ESCs, they select prices to the users. The users in the last stage, subscribe to one of the SAs.}
\label{fig:market}
\vspace{-0.3in}
\end{figure}

We model the overall setting as a game with the SAs and the users as the players.
\begin{color}{black} Each SA's pay-off in this game is its profit (cf.~(\ref{eq:profit})), while each user's objective is the expected pay-off described in Section~\ref{sec:user}. 
This game consists of the following stages:
\begin{enumerate}
\item In the first stage, each SA selects one of the ESCs and pays $\tilde{p}_j$ $j=A,B$ or selects to stay out of the market.
\item In the second stage, SA $i$ selects its price $p_i$ knowing the decisions made in stage 1. 
\item In the last stage, given the first two stages' decisions, the subscribers will choose one of the SAs from which to receive serve or choose not to receive service. We seek to characterize {\em Wardrop equilibrium} which is a notion used for non-atomic users. 
\end{enumerate}
%Knowing the first stage choices, in the second stage each SA $i$ will select a price $p_i$ to attract subscribers. If a SA does not obtain information from the ESCs, then, it can not offer any service to the customers. 
  We refer to a sub-game perfect Nash equilibrium of this game as a {\it market equilibrium}.
 Figure ~\ref{fig:market} depicts the tiered market architecture. 
   \end{color}
%\subsection{Stage Game}
%We model the market as three stages. In the first stage, the SAs decide which ESC to choose from knowing their prices, $\tilde{p}_j$, and their quality of information, $q_j$. In the second stage,  given the ESC choices, each SA selects a price to the customers. In the third stage, the users select one of the SAs or neither depending on the prices of the SAs, the expected congestion cost from each of the SAs, and their own valuation $v$. We seek to obtain a sub-game perfect equilibrium of the overall game.  For the third stage equilibrium, since the users are non-atomic, we characterize their behavior via a Wardrop equilibrium~\cite{Wardrop52}. This means that if user's select both SA's then the expected user pay-offs from each SA should be equal and non-negative. If a SA serves no user while another SA does, then the former SA's expected pay-off must be no greater than that of the later. Finally, if some users are not served, then the expected pay-off of any SA must be either zero or negative.

%% file: monopoly.tex
\section{Monopoly Scenario}\label{sec:mon}
 We, first, focus 
on the scenario where in the first stage only SA $1$ contracts with an ESC $j$ ($j \in \{A,B\}$) while the SA $2$ stays out of the market, i.e., SA $2$'s payoff obtained from contracting with either ESC is negative.
Subsequently, we characterize the scenario where only SA $2$ obtains information from ESC $j$.  
%Note that if a SA's payoff by obtaining information from any of the ESCs leads to a negative profit, the SA will prefer not to obtain information from any one of the ESCs. Hence, the other SA will have a monopoly power in that scenario.

\subsection{Third Stage User Equilibrium}
Since SA $2$ is not in the market, the expected pay-off of SA $1$'s users is simply given by averaging over (\ref{eq:cong_license}) and (\ref{eq:cong_diff}), i.e., 
\begin{align}\label{eq:monuser_pay}
q_jv-q_j\dfrac{\alpha^2\lambda_1}{W-L}-q_j\dfrac{(1-\alpha)^2\lambda_1}{L}-p_1.
\end{align}
Note here, we are averaging over both the probability a user is assigned to the licensed or unlicensed band as well as the probability that the spectrum is available.  It follows that since the users are non-atomic we have  $\lambda _1 = 0$ if 
$p_1 \geq q_jv$; otherwise, $\lambda_1$ is equal to the minimum of $\Lambda$ or the root of (\ref{eq:monuser_pay}).

%It can be seen that this quantity is maximized at a value between $0$ and $1$.

\subsection{Second Stage Pricing Strategy}

In the second stage with only SA $1$ in the market, it simply selects a price $p_1$ to maximize $p_1\lambda_1$, where $\lambda_1$ satisfies the third-stage equilibrium conditions above. 

\begin{theorem}\label{thm:monosa1}
The unique second stage pricing strategy is
\begin{align}
p_1=\min\{q_jv/2,q_jv-q_j\dfrac{\alpha^2\Lambda}{W-L}-q_j\dfrac{(1-\alpha)^2\Lambda}{L}\}.
\end{align}
The resulting third stage  equilibrium is
\begin{align}
\lambda_1=\min\{\dfrac{v/2}{\alpha^2/(W-L)+(1-\alpha)^2/L},\Lambda\}.
\end{align}
\end{theorem}
From (\ref{eq:monuser_pay}) and Theorem~\ref{thm:monosa1} we obtain the user's surplus is always zero in the monopoly scenario. Thus, if only SA $1$ obtains information from an ESC, the user's surplus is always zero. 

We also note that the payoff is not maximized when $\alpha=0$ or $\alpha=1$. Hence, if SA $1$ would have chosen $\alpha$, it would serve the users using the combination of licensed and unlicensed bandwidth. The payoff of SA $1$ also increases as $W-L$ increases, or $L$ increases. In Section~\ref{sec:first_stage}, we show that if $0<\alpha<1/2$, and $L$ is very large compared to $W-L$, the monopoly scenario arises where only SA $1$ obtains information from an ESC. 

%In Section~\ref{sec:first_stage} we show that if $0<\alpha<1/2$, and $L$ is very large compared to $W-L$, the monopoly scenario arises where only SA $1$ obtains information from ESC. 

\subsection{SA $2$ has a monopoly power}
Note from (\ref{eq:cong_diff2}) that when only SA $2$ obtains information from the ESC $j$, the expected payoffs of subscribers of SA $2$ are
\begin{align}
q_jv-q_j\dfrac{\lambda_2}{W-L}-p_2
\end{align}
The expected payoff of the subscribers is independent of $\alpha$ since SA $1$ is not in the market. 

Note that the pricing strategy in the monopoly scenario only depends on the congestion cost inflicted by own subscribers. Hence,
\begin{cor}
The monopoly pricing strategy for SA $2$, and the third stage equilibrium $\lambda_2$  is the same as the monopoly pricing strategy for SA $1$, and $\lambda_1$, respectively (stated in Theorem~\ref{thm:monosa1}) with $1/(W-L)$ in place of $\alpha^2/(W-L)+(1-\alpha)^2/L$. 
\end{cor}
Unlike the payoff of SA $1$, the payoff of SA $2$  only depends on $W-L$ and is independent of $\alpha$. The payoff of SA $2$ increases as $W-L$ increases.

Similar to the scenario where only SA $1$ is in the market, in this scenario, the users' surplus is also zero. 

%% file: same_esc.tex
\section{Both the SAs contract with the same ESC}\label{sec:same_esc}

 Next we consider the case where in stage 1, both SAs obtain information from ESC $j\in \{A,B\}$.
 
\subsection{Third Stage Equilibrium}
When both the SAs obtain information from the same ESC, they both always know that the spectrum is available at the same time. Thus, the ex-post payoff of the subscribers of SA $1$ in the unlicensed spectrum is given by (\ref{eq:same_cong}). By averaging over this and from
(\ref{eq:cong_license}), the expected pay-off of SA $1$'s users is given by
\begin{align}\label{eq:same_sa1}
\Pi_{1S}(\lambda_1,\lambda_2) = q_jv-\dfrac{q_j\alpha^2\lambda_1}{W-L}-\dfrac{q_j(1-\alpha)^2\lambda_1}{L}-\dfrac{q_j\alpha\lambda_2}{W-L}-p_1.
%q_jv-q_j\alpha g(\alpha(\alpha \lambda_1+\lambda_2)/(W-L))-q_j (1-\alpha)g((1-\alpha)\lambda_1/L)-p_1
\end{align}
Again we are averaging over both the spectrum availability ($q_j$) and the assignment of users to the two bands ($\alpha$). 

Since both SA's have the same information, the ex-post payoff of SA $2$'s subscribers is also given by (\ref{eq:same_cong}). Thus, the expected payoff of SA $2$'s users  is
\begin{align}\label{eq:same_sa2}
\Pi_{2S}(\lambda_1,\lambda_2) = q_jv-\dfrac{q_j\alpha\lambda_1}{W-L}-\dfrac{q_j\lambda_2}{W-L}-p_2.
%q_jv-q_j g\left(\dfrac{\alpha \lambda_1+\lambda_2}{W-L}\right)-p_2
\end{align}
Note that since SA $1$ serves the users with probability $\alpha$ in the unlicensed band, the terms in the congestion cost for SA $1$ due to the unlicensed band is multiplied by $\alpha$. Hence, this term is smaller than that of SA $2$. However, the  term due to the licensed band can be larger if $L$ is small compared to $W-L$. 

If both SA's serve traffic, then the users will be in a Wardrop equilibrium \cite{Wardrop52}, which specifies that 
\[
\Pi_{1S}(\lambda_1,\lambda_2)  = \Pi_{2S}(\lambda_1,\lambda_2) \geq 0.
\]
Examining the terms in this equation, it can be seen that more users will avail the service of SA $i$ if $v$ is large, and $p_i$ is small compared to $p_j$.
If on the other hand, SA $1$ does not serve traffic, then it must be that 
\[
\Pi_{1S}(0,\lambda_2) < \Pi_{2S}(0,\lambda_2)
\]
and the corresponding equation would hold if SA $2$ does not serve traffic.
Finally it must also be the case that if some users are not served ($\lambda_1 + \lambda_2 \leq \Lambda$) then 
\[
\max_i \Pi_{iS}(\lambda_1,\lambda_2) = 0.
\]
This last condition shows that  the users' expected payoffs are positive only when the the total number of users of both the SAs is equal to the total number of users present in the system,  $\Lambda$. Intuitively, users will avail service from the SAs as long as the expected payoff is positive. Thus, if the total number of users is less than $\Lambda$, the expected payoff must be zero.

Now, we state an observation which shows that if the prices of the SAs are the same, a scenario can not arise where the users only subscribe to SA $2$. 
\begin{lem}\label{lm:demand}
If $p_1=p_2$ and $\lambda_2>0$, then $\lambda_1>0$. 
\end{lem}
Intuitively, SA $1$ has a  competitive advantage over SA $2$ since SA $1$ can use both the unlicensed and licensed bands. However, SA $2$ can not use the licensed band. Thus, if the prices are the same, the congestion cost is smaller for SA $1$ compared to SA $2$ at least in the unlicensed band. Thus, the number of subscribers of SA $1$ can not be zero if $\lambda_2>0$. 
 
 %Also note that if $\alpha=1$, the expected payoff is the same for each of the SAs which leads to a price war. We will show that in this case, the price of the SAs must be zero. 

% Suppose that SA $1$  obtains information from an ESC 
%$j$, while the other SA chooses not to acquire information from either ESC (and so does not serve any customers).  The users of SA $i$ then obtain an expected pay-off of 
%\begin{align}
%q_jv-q_j\alpha g(\alpha \lambda_i/(W-L))-q_j(1-\alpha) g((1-\alpha) \lambda_i/L)-p_i.
%\end{align}
 \begin{color}{black}
\subsection{Second Stage Equilibrium}
Next we turn to the second stage pricing equilibrium. 
Our analysis reveals that
\begin{itemize}[leftmargin=*]
\item If $\alpha$ is above a threshold ($T$), the price and number of subscribers of SA $2$ is zero irrespective of the value of $v$ (Theorem~\ref{thm:opt}). Since SA $2$ has to incur a cost $\tilde{p}_j$ to acquire information from ESC $j$, its payoff is negative.  Thus, SA $2$ will be out of the market if there is a single ESC and $\alpha$ is high. 

When $\alpha$ is high, SA $1$ also serves users using the unlicensed spectrum with a higher probability. Thus, the congestion costs of both the SAs become similar which leads to a price war, i.e., the prices of both the SAs decrease. However, since SA $1$ also uses the licensed band, it still has an advantage.  Thus, though the price of SA $2$ is zero, the price of SA $1$ is still positive.  This shows that regulation may be required on the load a PAL holder puts on the unlicensed spectrum to maintain competition if there is a single ESC.%Hence, if $\alpha$ is high enough, the scenario where both the SAs will obtain information from the same ESC is not sustainable in an equilibrium path.
\item We also show that the threshold $T$ increases as $W-L$ decreases compared to $L$.  However, the threshold never reaches zero. Thus, if $W-L$ is small, the number of subscribers of SA $2$ goes to zero even for moderate values of $\alpha$.  Intuitively, if $W-L$ is small, the congestion cost of SA $2$ is also higher which reduces the number of subscribers of SA $2$. 

When $0<\alpha<1/2$ and $W-L$ is small compared to $L$, the SA $2$'s price also decreases to $0$. Thus, SA $1$ will have a monopoly power. Hence, if the ratio $\frac{W-L}{L}$ is small, SA $2$ will try not to obtain information from the same ESC as SA $1$. 
%\item We show that if $\dfrac{\alpha}{2(1-\alpha)}\geq \dfrac{W-L}{L}> \dfrac{2\alpha-1}{2(1-\alpha)}$, and $v$ is below a threshold, the number of subscribers of SA $2$ is zero (Theorem~\ref{thm:mode}). However, if $v$ is above the threshold, both the SAs have non-zero subscribers (Theorem~\ref{thm:mode}). Intuitively, higher $v$ indicates that the expected payoff of users will be higher. Thus, even though the congestion cost is higher, SA $2$ has a positive number of subscribers. 
\item Our analysis reveals that if $\frac{W-L}{L}>\frac{\alpha}{2(1-\alpha)}$, both the SAs select a positive price and have a positive number of subscribers (Theorem~\ref{thm:low}). Thus, if the ratio $\frac{W-L}{L}$ is high, and $\alpha$ is small, both the SAs will have positive number of subscribers when both the SAs obtain information from the same ESC. Note that if all the spectrum band is unlicensed as in \cite{icc_18}, the prices of both the SAs are zero when they obtain infor. However, we show that for any $L>\epsilon$, both the SAs will select a positive price if $\alpha<1$. %Our analysis also shows that if $v$ is below a threshold, the payoff of the SAs increase with both $L$, and $W-L$. However, when $v$ exceeds the threshold, the payoff decreases with $L$, and $W-L$. Thus, the amount of bandwidth only increases the payoff of the SA when the bandwidth is below a certain threshold. 
%\item We show that even when $0<\alpha<1/2$, and the ratio $\dfrac{W-L}{L}$ is very small, there exists an equilibrium where only SA $1$ has a positive payoff. Thus, if $L$ is very large compared to $W-L$, even for moderate values of $\alpha$, the SA $2$ may be out of the market. 
\item The payoff of SA $2$ increases as $\alpha$ decreases since higher $\alpha$ increases the congestion cost in the unlicensed spectrum which in turn reduces  SA $2$'s demand. However, surprisingly, SA $1$'s payoff does not always increase as $\alpha$ decreases. SA $1$'s payoff is always higher than that of the SA $2$'s. 

\item User's surplus is maximized neither at higher value of $W-L$ nor at higher a value of $L$. Further, the user's surplus is {\em neither} maximized at $\alpha=0$ (i.e., when both the SAs do not interfere with each other) nor at $\alpha=1$ . {\em Thus, the regulator has to again incentivize the SA $1$ to utilize the unlicensed band when there is a single ESC.}
%SA $1$'s payoff is higher than that of SA $2$ if $L$ is larger than $W-L$ since the congestion cost is lower when $W-L$ is smaller for SA $2$. 
\end{itemize}

\subsubsection{Results}
We now describe the pricing results in detail. We start with characterizing the scenario where the price and the number of subscribers of SA $2$ are zero if $\alpha$ is high. 
%We, now, characterize the equilibrium pricing strategies.. 

\begin{theorem}\label{thm:opt}
If $\frac{W-L}{L}\leq \frac{2\alpha-1}{2(1-\alpha)}$, 
%\begin{align}
%v\geq \alpha \Lambda/(W-L)
%%v\geq \dfrac{\alpha%\dfrac{\Lambda[\dfrac{\alpha-\alpha^2+2-2\alpha}{3L}+\dfrac{1-\alpha^2}{3(W-L)}]}{(1-\alpha)(1/L+1/(W-L))}+\Lambda(1-\alpha)[\dfrac{2-2\alpha}{3L}-\dfrac{2\alpha-1}{3(W-L)}]%\dfrac{\alpha^2\Lambda}{(W-L)}+\dfrac{(1-\alpha)^2\Lambda}{L}+\dfrac{\Lambda(1-\alpha)}{2(W-L)} 
%\end{align}
the unique second stage equilibrium pricing strategy is
\begin{align}
p_1&=q_j\min\{v,\dfrac{\alpha\Lambda}{W-L}\}\left(1-\dfrac{W-L}{\alpha}(\dfrac{\alpha^2}{W-L}+\dfrac{(1-\alpha)^2}{L})\right),\nonumber\\
p_2&=0.\nonumber
\end{align}
The third stage Wardrop equilibrium is
\begin{align}\label{eq:mon_lam}
\lambda_1=\min\{\dfrac{v(W-L)}{\alpha},\Lambda\}\quad \lambda_2=0.
\end{align}
%If $\alpha\geq \dfrac{2/L+1/(W-L)}{2/L+2/(W-L)}$, and $v\geq\alpha\Lambda/(W-L)$,
%the unique second stage equilibrium pricing strategy is
%\begin{align}\label{eq:mon_pr1}
%p_1=q_j\alpha\Lambda/(W-L)-q_j(\dfrac{\alpha^2\Lambda}{W-L}+\dfrac{(1-\alpha)^2\Lambda}{L})\nonumber\\
%p_2=0
%\end{align}
%The third stage wardrop equilibrium is
%\begin{align}\label{eq:mon_lam1}
%\lambda_1=\Lambda\nonumber\\
%\lambda_2=0.
%\end{align}
\end{theorem}
Note that the condition in Theorem~\ref{thm:opt} is clearly satisfied when $\alpha$ is equal to $1$. The condition $\frac{W-L}{L}\leq \frac{2\alpha-1}{2(1-\alpha)}$ is equivalent to the condition $\alpha\geq\dfrac{2/L+1/(W-L)}{2/L+2/(W-L)}$. Thus, if $W-L$ is smaller compared to $L$, the condition is more likely to be satisfied for a smaller value of $\alpha$.  The condition in the theorem is never satisfied when $\alpha<1/2$. 

The number of subscribers of SA $2$ is zero and only SA $1$ has a positive customer base.  Thus, the payoff of SA $2$ is negative in this scenario since SA $2$ still has to pay $\tilde{p}_j$ for obtaining information from ESC $j$.   
  However, SA $2$ would unilaterally deviate and may not obtain information from the ESC. In that scenario, its payoff or profit is zero. %Thus, the equilibrium where both the SAs obtain information from the same ESC is not sustainable when $\frac{W-L}{L}\leq \frac{2\alpha-1}{2(1-\alpha)}$.
  
  \begin{color}{black}
 {\em This result can be generalized to the case where there are multiple GAA tier SAs and they  obtain information from the same ESC. } The prices of these SAs must be zero since GAA tier SAs only use the unlicensed band. If there are more GAA tier SAs than ESCs, then multiple GAA tier SAs must obtain information from the same ESC. Hence, the prices of those GAA tier SAs have to be zero. Thus, those GAA tier SAs will not enter the market in the first place. Hence, in equilibrium the maximum number of GAA tier SAs must be less than or equal to the number of available ESCs. 
\end{color}

Also note that if $v\geq \alpha\Lambda/(W-L)$ and  $\frac{W-L}{L}\leq \dfrac{2\alpha-1}{2(1-\alpha)}$, SA $1$ obtains all the subscribers. Intuitively, if $v$ is large enough, the user's expected payoff increases, hence, more users will subscribe. The user's surplus is zero in this scenario. The SA $1$'s payoff is {\em not} maximized at $\alpha=0$ or at $\alpha=1$. 
We next introduce a parameter as a function of $\alpha$. We, later, show that if $v$ exceeds the parameter and $\alpha$ is not very high, none of the SAs have zero customers. 
\begin{definition}
Let 
\begin{align}\label{eq:positiveexpay}
\beta(\alpha)=&\dfrac{\Lambda}{W-L}\left[\dfrac{\dfrac{(2-\alpha-\alpha^2)(W-L)}{3L}+\dfrac{1-\alpha^2}{3}}{(1-\alpha)(\dfrac{W-L}{L}+1)}\right]\nonumber\\ & + \Lambda(1-\alpha)\left[(\dfrac{(2-2\alpha)(W-L)}{3L}-\dfrac{2\alpha-1}{3})\right].
%,\nonumber\\
%\dfrac{\Lambda[\dfrac{2-3\alpha+4\alpha^2-2\alpha^3}{3L(W-L)}+\dfrac{(1-\alpha)^3}{3L^2}+\dfrac{\alpha-\alpha^3}{3(W-L)^2}}{(1-\alpha)(1/L+1/(W-L))}+\Lambda(1-\alpha)[\dfrac{1-\alpha}{3L}+\dfrac{2-\alpha}{3(W-L)}]\}
\end{align}
\end{definition}
\begin{theorem}\label{thm:mode}
If $\frac{\alpha}{2(1-\alpha)}> \dfrac{W-L}{L}\geq\frac{2\alpha-1}{2(1-\alpha)}$, and $v\geq \beta(\alpha)$,
the second stage unique pricing strategy is
\begin{align}
p_1&=q_j\Lambda(1-\alpha)[\dfrac{1-\alpha}{3L}+\dfrac{2-\alpha}{3(W-L)}]\nonumber\\
p_2&=q_j\Lambda(1-\alpha)[\dfrac{2-2\alpha}{3L}-\dfrac{2\alpha-1}{3(W-L)}].
\end{align}
In the third stage equilibrium,
\begin{align}
\lambda_1&=\dfrac{p_1}{q_j(1-\alpha)^2(1/L+1/(W-L))}\nonumber\\
\lambda_2&=\dfrac{p_2}{q_j(1-\alpha)^2(1/L+1/(W-L))}.
\end{align} 
If $\frac{\alpha}{2(1-\alpha)}> \frac{W-L}{L}\geq\frac{2\alpha-1}{2(1-\alpha)}$, and $\beta(\alpha)>v$, the pricing strategy is the same as that in Theorem~\ref{thm:opt}.

%Further, if $v<\alpha\Lambda/(W-L)$, in the equilibrium, SA $2$ selects the price $0$. 
\end{theorem}
Note that $\beta(\alpha)$ is larger when  the ratio $\frac{W-L}{L}$ is either too small, or too large. The condition in Theorem~\ref{thm:mode} is less likely to occur if $\frac{W-L}{L}$ is either too small or too large.  Also note that if $v>\beta(\alpha)$,  $\lambda_1+\lambda_2=\Lambda$. Thus, all the users are served by the SAs. The user's surplus is also positive in this scenario.

%Note that the payoff of SA $1$ decreases with the increase in $\alpha$ when $v\geq \beta(\alpha)$. Thus, the SA $1$ would prefer the scenario where it will serve the users using the licensed band only.  Note that prices of both the SAs also decrease as $\alpha$ increases. The payoff of SA $1$ ($2$, resp.) is higher compared to SA $2$ ($1$, resp.) if $L$ is higher (lower, resp.) compared to $W-L$. 

Note from Theorem~\ref{thm:mode} that if $v<\beta(\alpha)$,  SA $2$'s price is zero in the equilibrium since the equilibrium price strategy becomes equal to the one stated in Theorem~\ref{thm:opt}. Hence, SA $2$ would not obtain the information from the ESC in the first place. Thus, the equilibrium where both the SAs obtain information from the same ESC is also not sustainable if $v<\beta(\alpha)$, and $\frac{\alpha}{2(1-\alpha)}> \frac{W-L}{L}\geq \frac{2\alpha-1}{2(1-\alpha)}$.

Note that if $W-L$ is small compared to $L$, the condition $v<\beta(\alpha)$ is likely to be achieved. Further, if $W-L$ is small compared to $L$, the condition $\frac{\alpha}{2(1-\alpha)}> \frac{W-L}{L}$ is also likely to be achieved even for smaller value of $\alpha$. Hence, if $W-L$ is small compared to $L$, the price of SA $2$ becomes zero and SA $1$ only achieves a monopoly.  
We, now, show that both the SAs select positive price, and have positive subscription base when $\alpha$ is small and $\frac{W-L}{L}$ is large. 
%We, first, introduce a parameter which is a function of $\alpha$. Let,
%\begin{align}%\label{eq:low}
%f(\alpha)= \dfrac{\dfrac{\Lambda}{L}(\dfrac{4(1-\alpha)^2(W-L)}{L}+3\alpha^2)}{\dfrac{2(1-\alpha)^2(W-L)^2}{L^2}+\dfrac{(2-2\alpha+3\alpha^2)(W-L)}{L}+\alpha^2}\nonumber
%\end{align}
%We show that if 
\begin{theorem}\label{thm:low}
If $\frac{W-L}{L}>\frac{\alpha}{2(1-\alpha)}$, the equilibrium is unique, and both SAs' second-stage prices are strictly positive. The user's surplus is positive only when $v>\beta(\alpha)$.
\end{theorem}
Thus, a small amount of licensed spectrum is sufficient for any $\alpha>0$ to achieve the condition that SA $2$ has a positive revenue when it obtains information from the same ESC as SA $1$. 

The user's surplus is positive only when $v>\beta(\alpha)$. Note from Theorem~\ref{thm:mode} that even when $\frac{\alpha}{2(1-\alpha)}> \dfrac{W-L}{L}\geq\frac{2\alpha-1}{2(1-\alpha)}$, and $v\geq \beta(\alpha)$, the user's surplus is positive. Thus, only when  $\dfrac{W-L}{L}\geq\frac{2\alpha-1}{2(1-\alpha)}$, and $v\geq \beta(\alpha)$, the user's surplus is positive. As we have mentioned, the inequality $v\geq \beta(\alpha)$ is satisfied only when the ratio between the unlicensed and licensed band is not too large nor too small. $\beta(\alpha)$ also decreases with $\alpha$. However, as $\alpha$  increases the inequality $\dfrac{W-L}{L}\geq\frac{2\alpha-1}{2(1-\alpha)}$ is less likely to be satisfied. Hence, the user's surplus is positive only when the ratio of unlicensed band and licensed band is of moderate values and $\alpha$ is neither too small nor too large. Intuitively, if $W-L$ is too small, the price of SA $2$ becomes zero and SA $1$ enjoys a monopoly power unless $\alpha=1$. On the other hand, if $L$ is small, the congestion cost is higher which decreases the user's surplus. %We now characterize the equilibrium when $\frac{W-L}{L}>\frac{\alpha}{2(1-\alpha)}$. 
\end{color}

%% file: diff_esc.tex
\section{SAs obtain information from different ESCs}\label{sec:diff_esc}
Next we turn to the case where in stage 1, the SAs obtain information from different ESCs.
We, first, consider the scenario where SA $1$ obtains information from ESC $A$, and SA $2$ obtains information from ESC $B$ (Section~\ref{sec:1a2b}). Subsequently, we consider the scenario where the SA $1$ obtains information from ESC $B$, and SA $2$ obtains information from ESC $A$ (Section~\ref{sec:2a1b}). %Throughout this section, we assume that $\tilde{p}_j$ is identical for both the ESCs. Note that $\tilde{p}_j$ is an exogenous parameter, and does not impact the pricing strategy of SAs (Second stage equilibrium), and the user's equilibrium (third stage equilibrium). However, it will impact the first stage equilibrium, {\em i.e.}, the selection of the ESC from which a SA will obtain information. The characterization of the equilibrium when $\tilde{p}_j$ is different is left for the future. 
\begin{color}{black}
\subsection{SA $1$ obtains information from ESC $A$}\label{sec:1a2b}
We, first, describe the third stage Wardrop equilibrium. We, subsequently, characterize the second stage price equilibrium. 
\subsubsection{Third stage Wardrop equilibrium}
 We, first, characterize the expected payoff of the users. Recall that when ESC $B$ estimates that the channel is available, the ESC $A$ also estimates that the channel is available. Thus,  the subscribers of SA $2$ always face congestion from the fraction of the SA $1$'s subscribers which are served using the unlicensed band.  However, the subscribers of SA $1$ that are served using the unlicensed band only face congestion from SA $2$'s subscribers when ESC $B$ indicates that  the incumbent is not present (which occurs with probability $q_B$).  Thus,  the subscribers of SA $1$ do not face congestion from the subscribers of SA $2$ in the unlicensed band with probability $q_A-q_B$.  Thus, from (\ref{eq:cong_license}), (\ref{eq:cong_diff}) and (\ref{eq:same_cong}), the expected payoff of the subscribers of SA $1$ is 
\begin{align}\label{eq:expay_sa1}
q_Av-(q_A-q_B)\dfrac{\alpha^2\lambda_1}{W-L}-\dfrac{q_B\alpha (\alpha\lambda_1+\lambda_2)}{W-L}-\dfrac{q_A(1-\alpha)^2\lambda_1}{L}-p_1.
\end{align} 
Note that the expected payoff of the subscribers of SA $1$ is not maximized when $\alpha=0$ or $\alpha=1$ rather at some value in between $0$ and $1$. 
%Thus, the users' expected pay-off in this scenario is% This results in these users having an expected pay-off of 
%\begin{align}
%q_Av-(q_A-q_B)\alpha g(\alpha\lambda_1/(W-L))-q_B\alpha g(\alpha\dfrac{\lambda_1+\lambda_2}{W-L})-(1-\alpha)q_Ag((1-\alpha)\lambda_1/(L/2))p_1.
%\end{align} 
%On the other hand, a user of SA $2$ will obtain an expected pay-off of 
%\begin{align}
%q_Bv-q_B\alpha g(\alpha\dfrac{\lambda_1+\lambda_2}{W-L})-(1-\alpha)g((1-\alpha)\lambda_2/(W-L))-p_2.
%\end{align} 
%
%Finally, suppose one SA $i$ $i\in \{1,2\}$ obtains information from an ESC 
%$k$, while the other SA chooses not to acquire information from either ESC (and so does not serve any customers).  The users of SA $i$ then obtain an expected pay-off of 
%\begin{align}
%q_kv-q_k\alpha g(\alpha \lambda_i/(W-L))-q_k(1-\alpha) g((1-\alpha) \lambda_i/(L/2))-p_i.
%\end{align}

%The  subscribers of SA $1$ only faces congestion from the subscribers in the unlicensed spectrum when ESC $B$ also states that the channel is available (it occurs w.p. $q_B$). 

Since when ESC $B$ informs that the channel is available, ESC $A$ also does the same, it follows  from (\ref{eq:same_cong})  that the expected payoff of the subscribers of $2$ is
\begin{align}\label{eq:expay_sa2}
q_Bv-q_B((\alpha \lambda_1+\lambda_2)/(W-L))-p_2.
\end{align}
The expected payoffs of subscribers of SA $2$ always decrease with $\alpha$. 

In a Wardrop equilibrium, the following must be true:  (i) If $\lambda_1>0$, then the expression in (\ref{eq:expay_sa1}) has to be non-negative and the expected payoff from SA $1$ has to be greater than or equal to that of SA $2$; (ii) if  $\lambda_2>0$ the expression in (\ref{eq:expay_sa2}) has to be non-negative and the expected payoff from SA $2$ has to be greater than or equal to that of SA $1$. 

If both $\lambda_1$ and $\lambda_2$ are positive, the expected payoff of the subscribers is again identical for both SA $1$ and SA $2$. It is evident that if $p_2>q_Bv$ $\lambda_2=0$; and if $p_1>q_Av$, $\lambda_1=0$. 

Lemma~\ref{lm:demand} also holds in this scenario. Thus, if $\lambda_2>0$, and $p_2=p_1$, $\lambda_1>0$. The above result will lead to the fact that SA $1$ will never select zero price. 

%\begin{align}
%& q_Av-(q_A-q_B)\alpha g(\alpha \lambda_1/(W-L))-q_A(1-\alpha)g((1-\alpha) \lambda/L)-q_B\alpha g((\alpha \lambda_1+\lambda_2)/(W-L))-p_1\geq 0\nonumber\\
%& q_Av-(q_A-q_B)\alpha g(\alpha \lambda_1/(W-L))-q_A(1-\alpha)g((1-\alpha) \lambda/L)-q_B\alpha g((\alpha \lambda_1+\lambda_2)/(W-L))-p_1\geq\nonumber\\ 
%& q_Bv-q_Bg((\alpha \lambda_1+\lambda_2)/(W-L))-p_2
%\end{align}
%If $\lambda_2>0$, then 
%\begin{align}
%& q_Bv-q_Bg((\alpha \lambda_1+\lambda_2)/(W-L))-p_2\geq 0\nonumber\\
%& q_Bv-q_Bg((\alpha \lambda_1+\lambda_2)/(W-L))-p_2\geq \nonumber\\
%& & q_Av-(q_A-q_B)\alpha g(\alpha \lambda_1/(W-L))-q_A(1-\alpha)g((1-\alpha) \lambda/L)-q_B\alpha g((\alpha \lambda_1+\lambda_2)/(W-L))-p_1
%\end{align}
%\end{lem}

\subsubsection{Second Stage Equilibrium}
%We assume that $g(\cdot)$ is linear. 
Our analysis reveals that--
\begin{itemize}[leftmargin=*]
%\item Unlike the scenario where both the SAs obtain information from the same ESC, in this scenario, both the SAs have positive subscription base even when $\alpha$ is large if $v$ is below a threshold ($T$).  Thus, even though SA $2$ has an inferior quality of information, the payoff of SA $2$ is not zero when $\alpha$ is large.  
%\item The threshold $T$ is a decreasing function of $L$, and $W-L$. Thus, if both the amount of licensed and unlicensed bandwidth are higher, SAs will more unlikely to opt for obtaining information from different ESCs. Thus, if the amount of licensed and unlicensed bandwidth are high, SAs will compete for the same ESC. 
\item Unlike the scenario where both the SAs obtain information from the same ESC, if the ratio $\eta = \frac{W-L}{L}$ is small and $\alpha$ is high (Corollary~\ref{cor:highalpha}),  SA $2$ sets a positive price. 
\item Unlike the scenario where both the SAs obtain information from the same ESC, if $v$ is large, the profit of SA $2$ is negative. %Hence, if $v$ is large, the equilibrium where SA $1$ obtains information from ESC $A$, and SA $2$ obtains information from ESC $B$ is not sustainable.  
%We show that if $\alpha$ is below a critical value $\alpha_c$, the equilibrium where SA $1$ obtains information from ESC $A$, and SA $2$ obtains information from ESC $B$ is not sustainable as SA $2$ has an incentive to deviate and obtain information from ESC $B$. SA $2$ has a strictly higher payoff in this scenario compared to the one where both the SAs obtain information from the same ESC if $W-L$ is smaller than $L$.
%\item The payoff of SA $2$ is not necessarily maximized at $\alpha=0$ unlike the scenario where both the SAs obtain information from the same ESC. In fact, if $W-L$ is smaller than $L$, and $q_1<q_B<q_2<q_A$, the payoff is higher when $\alpha=1$. Popular perception indicates that the SA $2$'s payoff should decrease as $\alpha$ increases. However, if the SAs obtain information from different ESCs, the expected congestion cost can be reduced by obtaining information from a different ESC. Thus, the payoff of SA $2$ may increase when $\alpha=1$.
%\item There is a unique $\alpha$ which maximizes the SA $1$'s payoff. The maximum can also be reached at $\alpha=1$ if$W-L$ is smaller than $L$. 
\item Similar to the scenario where both the SAs obtain information from the same ESC, in this scenario there exists a monopoly equilibrium where only SA $1$ can serve a positive number of users if $0<\alpha<1/2$, and $W-L$ is very small compared to $L$. Intuitively, when $\alpha$ is small and $W-L$ is small, the users' expected payoffs are way better for SA $1$ compared to SA $2$.
%Intuitively, when $\alpha<1/2$, 
\end{itemize}

We, now, describe the results in detail. %First, we show that $v$ is below a threshold, both the SAs may have positive profit.
\begin{theorem}\label{thm:1a2b}
The second stage pricing strategy is unique. If 
\begin{align}\label{eq:condn1ahighv}
(q_A-q_B)v\leq &\dfrac{\Lambda}{W-L}(2q_A\alpha^2-3q_B\alpha+q_B)
+\dfrac{\Lambda}{L}(2q_A(1-\alpha)^2)
\end{align}
and 
\begin{align}\label{ineq:pos}
 \frac{W-L}{L}\geq \frac{q_B\alpha^2/q_A+\alpha-2\alpha^2}{2(1-\alpha)^2},
\end{align}
 then SA $2$ selects a positive price in the equilibrium. 

Otherwise, $p_2=0$ in the unique equilibrium.
%Consider the following second stage pricing strategy profile
%\begin{align}\label{eq:1a2bpr}
%p_1=\dfrac{(q_A-q_B)v}{3}+\dfrac{(q_A\alpha^2-3q_B\alpha+2q_B)\Lambda}{3(W-L)}+\dfrac{q_A(1-\alpha)^2\Lambda}{3L}\nonumber\\
%p_2=\dfrac{(q_B-q_A)v}{3}+\dfrac{(2q_A\alpha^2-3q_B\alpha+q_B)\Lambda}{3(W-L))}+\dfrac{2q_A(1-\alpha)^2\Lambda}{3L}
%\end{align}
%Under the second stage pricing strategy profile, the third stage Wardrop equilibrium is
%\begin{align}
%\lambda_1=\dfrac{p_1}{\dfrac{q_A\alpha^2}{W-L}+\dfrac{q_A(1-\alpha)^2}{L}+\dfrac{q_B(1-2\alpha)}{W-L}}\nonumber\\
%\lambda_2=\dfrac{p_2}{\dfrac{q_A\alpha^2}{W-L}+\dfrac{q_A(1-\alpha)^2}{L}+\dfrac{q_B(1-2\alpha)}{W-L}}
%\end{align}
%The pricing strategy stated in (\ref{eq:1a2bpr}) is a Nash equilibrium if 
%
%\begin{align}\label{eq:condn1a2b}
%q_Bv\geq p_2+q_B\alpha\lambda_1/(W-L)+q_B\lambda_2/(W-L)
%\end{align}
\end{theorem}
The complete characterization of the pricing strategy is omitted owing to the space constraint. 
The condition in (\ref{eq:condn1ahighv}) indicates that If the difference between $q_A$ and $q_B$ is high, the price of SA $2$ may be zero. Intuitively, if SA $1$ has a far superior information, SA $2$ must select a lower price. Also note that if $v$ is high, the price of SA $2$ can also be zero. If the valuation $v$ is high, the expected payoff achieved by the users is way better from SA $1$ compared to SA $2$.  Thus, SA $2$ has to select a lower price.  Hence, if the valuation of users is high, there may not be any competition between the SAs in this scenario. 

 Note that the  inequality  in  (\ref{eq:condn1ahighv})  is more likely to be satisfied when either $W-L$ or $L$ is small. Thus, if $W-L$ is small, or large compared to $L$, the price of SA $2$ will  be likely to be non-zero. Also 
Note that the prices of the SAs do not go to zero as $\alpha$ increases to $1$ unlike the scenario when SAs obtain information from the same ESC if the inequality in (\ref{eq:condn1ahighv}) is satisfied. 

Note from Theorem~\ref{thm:1a2b} that if $\frac{W-L}{L}\leq \frac{q_B\alpha^2/q_A+\alpha-2\alpha^2}{2(1-\alpha)^2}$, the price of SA $2$ is $0$. If $q_B$ increases, the price of SA $2$ is zero for a larger range of values of the ratio $\frac{W-L}{L}$.

When $\alpha> \dfrac{1}{2-q_B/q_A}$ , the right hand side of the inequality in (\ref{ineq:pos}) becomes negative, hence, the inequality is always satisfied. Note also that from Theorem ~\ref{thm:opt} that if $\frac{W-L}{L}\leq \dfrac{2\alpha-1}{2(1-\alpha)}$ the SA $2$'s price becomes zero if both the SAs obtain information from the same ESC. Thus, SA $2$'s price does not become zero when $\alpha$ is high unlike the scenario where the SAs obtain information from the same ESC. %Intuitively, when SAs obtain information from different ESCs, the competition becomes more relaxed because of the difference in the congestion cost even when $\alpha$ is high, thus, the price of SA $2$ can be positive. 
 
 %Also note that the inequality is not satisfied when $\alpha=1$, thus, the price of SA $2$ does not go to zero as $\alpha$ increases. 

%\begin{corollary}\label{cor:highalpha}
%If $\dfrac{q_B\alpha^2/q_A+\alpha-2\alpha^2}{2(1-\alpha)^2}<\dfrac{W-L}{L}\leq \dfrac{2\alpha-1}{2(1-\alpha)}$, and the condition in (\ref{eq:condn1ahighv}) is satisfied, SA $2$ will have a strictly higher payoff if SA $2$ obtains information from ESC $B$ rather than ESC $A$.
%\end{corollary}
%The condition in (\ref{eq:condn1ahighv}) is satisfied when $W-L$ is small compared to $L$. Also note that 
In the following, we evaluate the threshold above which SA $2$ will prefer to obtain inferior quality information compared to SA $1$.  Let $\alpha_c$ be the lowest possible positive value such that if $\alpha\geq \alpha_c$, the inequality in (\ref{ineq:pos}) is satisfied. $\alpha_c$ is a function of the ratio $\eta$. If $\eta$ is high, $\alpha_c$ is lower. Thus, 
%Note that
%\begin{align}
%\dfrac{q_B\alpha^2/q_A+\alpha-2\alpha^2}{2(1-\alpha)^2}< \dfrac{2\alpha-1}{2(1-\alpha)}
%\end{align}
%if $\alpha> \alpha_c$. Hence, 
\begin{corollary}\label{cor:highalpha}
If $\alpha\in[\max\{\alpha_c,\dfrac{2/L+1/(W-L)}{2/L+2/(W-L)}\}, 1]$ and the condition in (\ref{eq:condn1ahighv}) is satisfied SA $2$ will have a strictly higher payoff when it obtains information from ESC $B$ compared to the scenario where both the SAs obtain information from ESC $A$.
\end{corollary}
The inequality in (\ref{eq:condn1ahighv}) is achieved when $W-L$ is small. Thus, if $W-L$ is small compared to $L$, and $\alpha$ is high, the best response of SA $2$ is to obtain information ESC $B$ if SA $1$ obtains information from ESC $A$. Note that for other values of parameters, SA $2$ may still have an incentive to obtain information from ESC $B$ rather than from ESC $A$.

\subsection{SA $1$ obtains information from ESC $B$}\label{sec:2a1b}
We now consider the scenario where the SA $1$ obtains information from ESC $B$ and SA $2$ obtains information from ESC $A$.  

\subsubsection{Third Stage Wardrop Equilibrium} 
%Subsequently, we characterize the Wardrop equilibrium. 
Since SA $1$ now obtains information from ESC $B$, the fraction of the subscribers of SA $1$ served in the unlicensed spectrum now always face congestion from the subscribers of SA $2$. Thus, from (\ref{eq:same_cong}) the expected payoffs of the subscribers of SA $1$ are
\begin{align}\label{eq:sa1b}
q_Bv-q_B\alpha(\alpha\lambda_1+\lambda_2)/(W-L)-q_B(1-\alpha)^2\lambda_1/L-p_1.
\end{align}
 Note that the  subscribers of SA $2$ only faces congestion from the $\alpha\lambda_1$ subscribers of SA $1$ when ESC $B$ informs that the channel is available (which occurs with probability $q_B$).  However, the subscribers of SA $2$ do not face congestion from the subscribers of SA $1$ with probability.  $q_A-q_B$. Thus, The expected payoffs of the subscribers of SA $2$ are
\begin{align}\label{eq:sa2a}
& q_Av-q_B\dfrac{\alpha\lambda_1+\lambda_2}{W-L}-(q_A-q_B)\dfrac{\lambda_2}{W-L}-p_2\nonumber\\
& =q_Av-q_B\dfrac{\alpha\lambda_1}{W-L}-q_A\dfrac{\lambda_2}{W-L}-p_2
\end{align}

In the Wardrop equilibrium we then have: (i) $\lambda_1>0$,  if the expression in (\ref{eq:sa1b}) is non-negative, and the expected payoff attained by the users from SA $1$ is greater than or equal to the expected payoff attained from SA $2$ (cf.(\ref{eq:sa2a})); (ii) $\lambda_2>0$ if the expression in (\ref{eq:sa2a}) is non-negative, and the expected payoff attained by the users from SA $2$ is greater than or equal to the expected payoff attained from SA $1$ (cf.(\ref{eq:sa1b})).
%\begin{align}
%q_Bv-q_B\alpha g(\alpha\lambda_1+\lambda_2)-q_B(1-\alpha)g((1-\alpha)\lambda_1)-p_1\geq 0\nonumber\\
%q_Bv-q_B\alpha g(\alpha\lambda_1+\lambda_2)-q_B(1-\alpha)g((1-\alpha)\lambda_1)-p_1\geq\nonumber\\
%q_Av-q_Bg(\alpha\lambda_1+\lambda_2)-(q_A-q_B)g(\lambda_2)-p_2
%\end{align}
%$\lambda_2>0$, if
%\begin{align}
%q_Av-q_Bg(\alpha\lambda_1+\lambda_2)-(q_A-q_B)g(\lambda_2)-p_2\geq 0\nonumber\\
%q_Av-q_Bg(\alpha\lambda_1+\lambda_2)-(q_A-q_B)g(\lambda_2)-p_2\geq \nonumber\\
%q_Bv-q_B\alpha g(\alpha\lambda_1+\lambda_2)-q_B(1-\alpha)g((1-\alpha)\lambda_1)-p_1
%\end{align}
%When both $\lambda_1>0$, and $\lambda_2>0$, the expected payoff of subscribers of SAs obtain the same expected payoff. 
%Lemma~\ref{lm:demand} may not hold in this scenario. Thus, even if $p_1=p_2$, it does not imply that $\lambda_1>0$ when $\lambda_2>0$.

\subsubsection{Second Stage Equilibrium}
We, now, describe the second stage equilibrium. Our analysis reveals the following
\begin{itemize}[leftmargin=*]
\item Unlike the scenarios where both the SAs obtain information from the same ESC or SA $1$ has a superior information, in this scenario, SA $1$'s price goes to zero when $(q_A-q_B)v$ is above a threshold $T_1$ (Theorem~\ref{thm:1b2a}).  Note that since SA $1$ has to incur a cost for obtaining information from the ESC, hence, SA $1$'s profit is negative. Thus, such a scenario is not sustainable in the equilibrium. 
\item The Nash equilibrium is unique.  Both the SAs serve positive number of users if $\alpha$ is high and $v$ is below a threshold. Thus, unlike the scenario where both the SAs obtain information from the same ESC, the price of SA $2$ is positive when $\alpha$ is high. %If $v$ is above the threshold $T_1$, the price of SA $1$ goes to zero. Thus, if the user's valuation is large, there is no competition between the SAs when they obtain information from different ESCs. %However, note that when both the SAs obtain information from the same ESC. There is no equilibrium where SA $1$'s price becomes zero. Thus, SA $1$ would have an incentive to deviate and obtain information from ESC $A$ in the first stage. Hence, if $v$ is high, th4e equilibrium where the SA $1$ obtains information from ESC $B$, and SA $2$ obtains information from ESC $A$ is not sustainable when $v$ is high. %Thus, if $v$ is below the threshold $T_1$, there is no equilibrium where SA $1$ obtains information from ESC $B$, and SA $2$ obtains information from ESC $A$.
%\item The SA $1$ does not have any incentive to deviate for $\alpha<1/2$, if $L$ is large compared to $W-L$, and $L$ is small compared to $W-L$ if $\alpha\geq 1/2$. Intuitively, when $\alpha$ is small, the SA $1$ uses licensed band for serving the users more. Thus, only if $L$ is large the congestion cost will be lower. Thus, the equilibrium where SA $1$ has an inferior quality of information is sustainable when $L$ is large and $\alpha$ is small. On the other hand, if $\alpha$ is large, SA $1$ serves users more using the unlicensed band, Thus, the equilibrium where SA $1$ has an inferior quality of information is sustainable only when $W-L$ is large compared to $L$.
%\item There exists a critical value $\alpha_c$ such that if $\alpha\geq \alpha_c$, the equilibrium where SA $1$ obtains information from ESC $B$, and SA $2$ obtains information from ESC $A$ is sustainable. 
\item If $\alpha$ is small, and $W-L$ is small compared to $L$, the monopoly scenario may arise where only SA $1$ serves the users similar to the previous scenarios. However, unlike the previous scenarios, if $q_B$ is small, the monopoly scenario arises for a smaller range of values of $\alpha$. Intuitively,  if $q_B$ is smaller, SA $1$'s market power decreases since the users' expected payoff from SA $1$ decrease.  %\item If $\alpha$ is high, and $W-L$ is small, unlike the scenario where both the SAs obtain information from the same ESC, the prices of both the SAs are positive. 

%\item Similar to the scenario where both the SAs obtain information from the same ESC, in this scenario, if $0<\alpha<1/2$, and $W-L$ is very small compared to $L$, there exists an equilibrium where SA $2$ has zero subscription base. 
%\item Unlike the scenario where the SA $1$ obtains information from ESC $A$, and SA $2$ obtains information from ESC $B$, if $L$ is very large compared to $W-L$, the equilibrium where SA $1$ obtains information from ESC $B$, and SA $2$ obtains information from ESC $A$ is not sustainable in the equilibrium path. 
\end{itemize}
We, now, detail the results.%We, again, assume that $g(\cdot)$ is linear. 
\begin{theorem}\label{thm:1b2a}
The Nash equilibrium is unique. If 
\begin{align}\label{eq:pricehighv1b}
(q_A-q_B)v\leq\dfrac{(2q_A-3q_B\alpha+q_B\alpha^2)\Lambda}{3(W-L)}+\dfrac{q_B(1-\alpha)^2\Lambda}{3L};
\end{align}
and 
\begin{align}\label{ineq:pos2}
\dfrac{W-L}{L}\geq \dfrac{q_B\alpha^2+q_B\alpha-2q_A\alpha^2}{2q_A(1-\alpha)^2}
\end{align}
the pricing strategy of both the SAs are positive. 

If the inequality in (\ref{eq:pricehighv1b}) is not satisfied, the equilibrium price of SA $1$ becomes zero and if the inequality in (\ref{ineq:pos2}) is not satisfied, the equilibrium price of SA $2$ becomes zero.
\end{theorem}
Owing to the space constraint, we omit the complete characterization of the Nash Equilibrium. Note from (\ref{eq:pricehighv1b}) that if differences in the expected valuation (i.e., $(q_A-q_B)v$) is large, the price of SA $1$ may become zero. This is because, users will achieve a higher expected payoff from SA $2$ because of the superior information. 

%The inequality in (\ref{eq:pricehighv1b}) is satisfied when either $W-L$ is small or $L$ is small. Intuitively, if $W-L$ is too small compared to $L$, the price set by SA $2$ also becomes 

Note that when both the SAs obtain information from the same ESC. There is no equilibrium where SA $1$'s price becomes zero. Thus, SA $1$ would have an incentive to deviate and obtain information from ESC $A$ in the first stage if $v$ is large. Hence, if $v$ is large, the equilibrium where the SA $1$ obtains information from ESC $B$, and SA $2$ obtains information from ESC $A$ is not sustainable.

%We, now, identify the condition under which the SA $2$'s payoff may be positive, however, the SA $2$'s payoff is negative if it would obtain the information from ESC $B$ (same as SA $1$). 
Note from (\ref{ineq:pos2}), if $\alpha\geq \frac{1}{3-2q_B/q_A}$, the right hand side of the inequality becomes negative. Hence at least when $\alpha\geq \frac{1}{3-2q_B/q_A}$, the inequality in (\ref{ineq:pos2}) is satisfied.  Since $q_B<q_A$, hence, the price of SA $2$ is positive for high enough $\alpha$. Note that the inequality in (\ref{ineq:pos2})  is less likely to be satisfied if $W-L$ is small compared to $L$ and $q_B$ is small.

Note from Theorem~\ref{thm:opt} that if $\dfrac{W-L}{L}\leq \dfrac{2\alpha-1}{2(1-\alpha)}$, the price of SA $2$ is zero when both the SAs obtain information from the same ESC. Thus, unlike the scenario where both the SAs obtain information from both the ESCs, the SA $2$'s price is positive when $\alpha$ is high. 
\end{color}

%% file: first_stage.tex
\section{First Stage Decision}\label{sec:first_stage}
We now turn to the first stage in which the SA's decide on 
which ESC, if any, to contract with.  We first introduce some notation.

\begin{definition}
Let $\pi_i(j_1,j_2)$ denote the payoff of SA $i, i=1,2$ when SA $i$ obtains information from $j_i\in\{A,B,\phi\}$, where $\phi$ denotes that the SA $i$ did not select any ESC. 
\end{definition}

Using the above notation,  a Nash equilibrium profile in the first stage is defined as follows:

\begin{definition}\label{defn:nash}
A pair of choices $(j_1,j_2)$ is a Nash equilibrium profile if $\pi_1(j_1,j_2)\geq \pi_1(k_1,j_2)$, and $\pi_2(j_1,j_2)\geq \pi_2(j_1,k_2)$ for any  $k_i\in\{A,B,\phi\}, k_i\neq j_i$. 
\end{definition}

We have already characterized the second stage equilibrium pricing strategy, and the third stage Wardrop equilibrium for each of the possible first stage decisions. From Definition~\ref{defn:nash}, in order to conclude whether a strategy profile is a Nash equilibrium we need to compute the payoff of each possible decision of a SA, keeping the decision of the other SA fixed. If both the SAs can not attain a higher payoff by unilaterally deviating from the strategy profile, the strategy profile is an NE.  
%In Section~\ref{sec:numerical} we characterize the Nash equilibrium first stage strategy profile in an example setting. 

In the following, we describe some of the key characteristics of the first stage Nash equilibrium. 
%For the ease of exposition, we assume that $\tilde{p}_j$ is the same for both the ESCs. Thus, there is no equilibrium where both the SAs obtain information from ESC $B$ since at least one of the SAs will have an incentive to deviate and obtains information from the ESC $A$. 
\begin{theorem}
If $\alpha=0$, there is no equilibrium where both the SAs obtain information from different ESCs. If $\alpha=1$, there is no equilibrium where both the SAs obtain information from the same ESC.
\end{theorem}

Thus, if $\alpha=0$, only one of the ESCs can exist and if its prices are low enough it may sell information to both SAs.
%Intuitively, if $\alpha=0$, the SAs split the entire frequency band, and the subscribers of both the SAs do not interfere with each other. Hence, the congestion cost of a SA does not depend on the subscribers of other SA. Thus, if the SA which offers an inferior service will attain a lower payoff. 
On the other hand, if $\alpha=1$, in an equilibrium, the SAs may obtain information from different ESCs. When $\alpha=1$ and the SAs obtain information from the same ESC, then the congestion cost of both the SAs is identical irrespective of the subscribers of each SAs. This leads to a price war. Obtaining information from different ESCs enables the SAs to differentiate themselves and reduces the likelihood of a price war.
 %If they obtain information from the same ESC, the congestion cost is  

We, now, characterize the equilibrium when the ratio $\eta := \tfrac{W-L}{L}$ goes to $\infty$. %First, we characterize the scenario where both the 
\begin{theorem}\label{thm:lowl}
If $\tilde{p}_A$ is small enough, and the ratio $\eta \rightarrow \infty$, both the SAs obtain information from ESC $A$ in an equilibrium if $\alpha<1$.
\end{theorem}
 Note in \cite{icc_18} that there is no equilibrium where both the SAs obtain information from the same ESC if $\alpha=1$ even when $\tilde{p}_j$ is small. {\em Surprisingly, we show that even a small amount of licensed band will force the SAs to obtain information from the same ESC.} Intuitively, when $W-L$ is very large compared to $L$, the congestion cost of SA $1$ is high compared to SA $2$ for $\alpha<1$. Since the congestion cost in the licensed band becomes dominant factor, thus, the expected payoff of users are quite different even when the SAs obtain information from the same ESC. 
  
  Since $\tilde{p}_A$ is small, thus, SA $1$ would obtain information from ESC $A$ rather than from ESC $B$.  However, SA $1$'s payoff decreases as $L$ decreases. Hence, if $\tilde{p}_A$ is fixed, and positive, for small enough $L$, SA $1$ would not obtain information from ESC $A$. %If $\tilde{p}_B$ is also fixed, the SA $2$ would have a monopoly where SA $1$ does not obtain any information. 
% Thus, SA $1$ would have a smaller payoff compared to SA $2$ if they obtain information from the same ESC. SA $2$ would like to obtain information from ESC $A$. If SA $1$  obtains an information different from SA $2$, it would reduce the payoff of SA $1$. 
%
%On the other hand, if ESC $A$ has a higher price, SA $2$ would obtain information from ESC $B$. Even if SA $1$ obtains an information from ESC $A$, the payoff will still be smaller compared to the price $\tilde{p}_a$ because of the higher congestion cost of ESC $A$. 
%Thus, when $W-L$ is large compared to $L$, SAs obtain information from the same ESC. Note that when $W-L$ is large, the congestion cost in negligible for SA $2$ as well as for SA $1$. SA $2$ would still can select a relatively higher price and obtains a positive number of subscribers. Thus, the SAs would prefer superior quality of information as it would increase the payoff. 

We, next, consider the scenario where the ratio $\eta$ goes to zero. 
%The next theorem characterizes the condition under which a monopoly scenario may arise. If 
%\begin{theorem}
%If 
%\begin{align}
%\dfrac{\alpha}{W-L}>\dfrac{2\alpha^2}{W-L}+\dfrac{2(1-\alpha)^2}{L}
%\end{align}
%a monopoly scenario arises where only SA $1$ obtains information from the ESC $A$. 
%\end{theorem}
%When $W-L$ is small compared to $L$, the congestion cost is higher for SA $2$. When $\alpha$ is large, the above condition is not satisfied. When $\alpha$ increases, the congestion cost of SA $1$ also increases, thus, the price of SA $1$ also has to decrease. Hence, the monopoly scenario arises only when $0<\alpha<1/2$ and $W-L$ is small compared to $L$.

\begin{theorem}\label{thm:lowu}
If $0\leq\alpha\leq \frac{1}{2}$, and the ratio $\eta\rightarrow 0$, the monopoly scenario arises where only SA $1$ obtains information from one of the ESCs. %$L\rightarrow \infty$, and $W-L$ is finite, only when $\alpha\geq 1/2$, and $v\leq T$, there exists an equilibrium where SA $1$ obtains information from ESC $A$, and SA $2$ obtains information from ESC $B$.

If $\alpha>\frac{1}{2}$, and the ratio $\eta \rightarrow 0$, and $\tilde{p}_A, \tilde{p}_B$ ($\tilde{p}_A>\tilde{p}_B$) are small enough, there is an equilibrium where SA $1$ obtains information from ESC $A$, and SA $2$ obtains information from ESC $B$.
%If $\alpha<1/2$ or $v>T$, a monopoly scenario exists where SA $1$ obtains information from ESC $A$.
\end{theorem}
Thus, if the licensed bandwidth is very large compared to the unlicensed bandwidth, the monopoly scenario arises where only SA $1$ exists in the market if $\alpha$ is small enough. However, if $\alpha$ is larger, there can be an equilibrium where SA $1$ obtains information from ESC $A$, and SA $2$ obtains information from ESC $B$. Hence, both the ESCs can co-exist if $\alpha$ is large enough, and the ratio $\eta$ is small enough.

Intuitively, when $W-L$ is small compared to $L$, and $\alpha$ is small, the congestion cost faced by the subscribers of SA $2$ is large compared to SA $1$. Hence, SA $1$ has a  monopoly power. However, when $\alpha$ is large, the congestion cost faced by the subscribers of SA $1$ also becomes higher. Thus, the SA $2$ can exist in the market albeit by obtaining an inferior quality of information. Thus, multiple ESCs can co-exist only if the ratio between the licensed bandwidth and unlicensed bandwidth is high enough and $\alpha$ is high.

%% file: numerical.tex
\section{Numerical Evaluation}\label{sec:numerical}
In this section, we numerically compare the profits of SAs, the user surplus, and the social welfare as a function of $\alpha$ and the ratio $\eta = \tfrac{W-L}{L}$. This will enable us to determine the value of $\alpha$ and $L$ which should be set in order to achieve a given objective. 

We set $W = 150$ MHz. Note that this is consistent with the current CBRS proposal\cite{FCC-CBRS}.  We also set $v$ at $10$, $q_A$ at $0.6$, and $q_B$ at $0.4$; $\tilde{p}_A$ and $\tilde{p}_B$ are set at $1$, and $0.5$ respectively.  %We vary $L$ from $10$ Mhz to $140$ MHz. In other words, we 

\begin{figure}
%\begin{minipage}{0\textwidth}
\includegraphics[trim=0in 0in 0in 0in,width=0.4\textwidth]{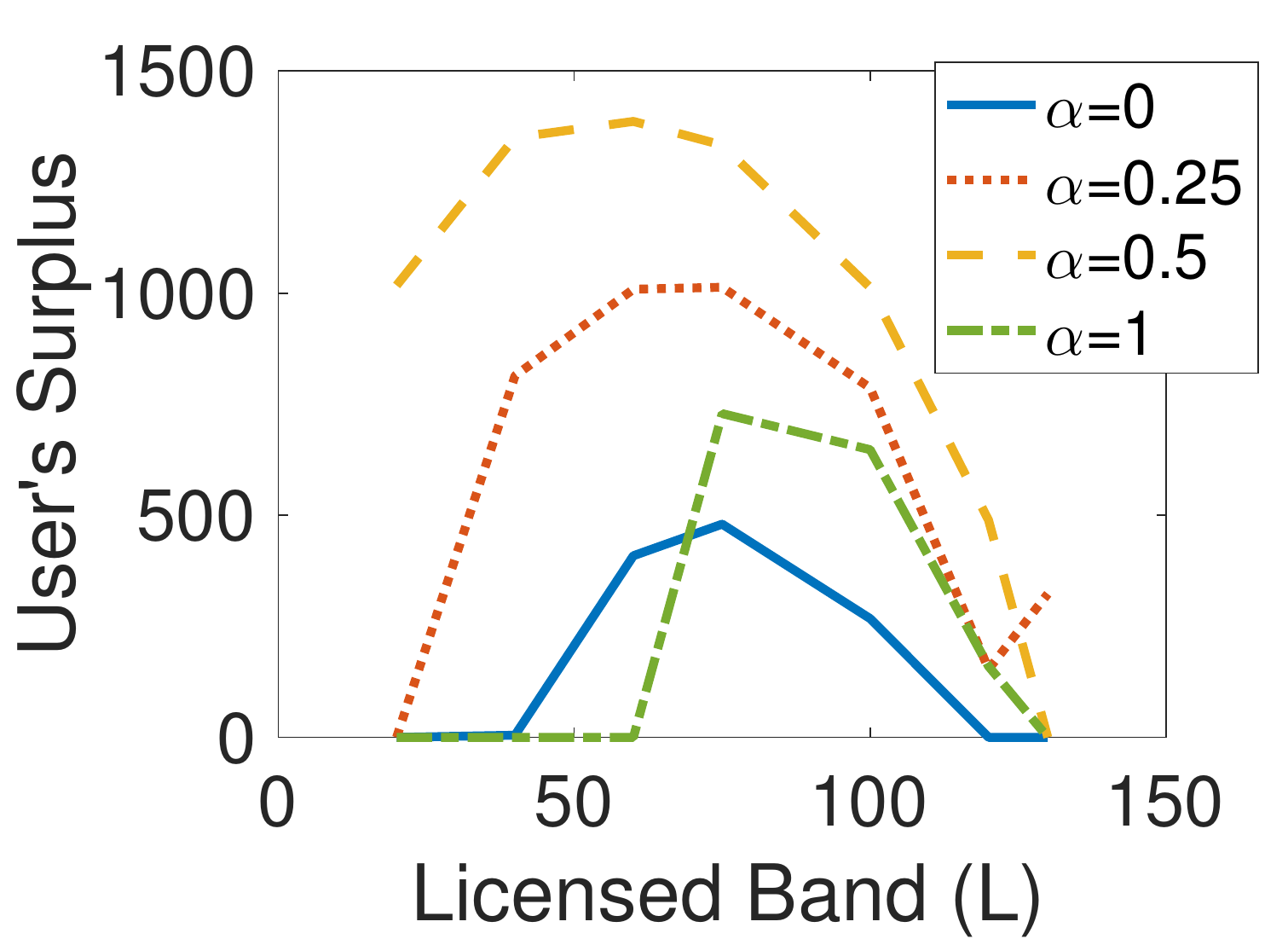}
\vspace{-0.1in}
\caption{The user surplus versus $L$ for different values of $\alpha$.}
\label{fig:surplus}
\vspace{-0.2in}
\end{figure}

\subsection{User's Surplus}
Figure~\ref{fig:surplus} shows that surprisingly the user's surplus is not maximum when $W-L$ achieves its largest value for any $\alpha$. The user's surplus increases as $L$ initially increases;  however, it decreases when $L$ exceeds a threshold. This contradicts the intuition that more unlicensed spectrum will always benefit user surplus.

When $L$ is small, the congestion cost is high for SA $1$. In this regime, both the SAs obtain information from the same ESC and serve a positive number of subscribers if $\alpha<1$. Thus, if $\alpha<1$, the higher congestion cost in the licensed band drives down the user's surplus.  When $\alpha=1$ the SAs obtain information from different ESCs. However, for smaller values of $L$, a monopoly arises where only one of the SAs obtain information from the ESCs. Hence, the user's surplus becomes small. 

On the other hand, if $L$ is large, the congestion cost of SA $2$ is higher. Thus, SA $1$ has a competitive advantage when $\alpha$ is small, enabling it to select a higher price. When $\alpha$ is high,  SA $1$'s congestion cost also increases since the amount of unlicensed bandwidth is small. Thus, the user's surplus also decreases. 

When the ratio $\eta$ is neither too high, nor too small, the total number of subscribers served by the SAs become equal to $\Lambda$. Hence, the user's surplus becomes positive. Note that we have already shown that if the ratio $\dfrac{W-L}{L}$ is moderate, the user's surplus is the highest. This is in accordance with our theoretical results. 

When $\alpha$ increases, the user surplus increases. This is because as $\alpha$ increases, the competition between the SAs become more intense. However, the user's surplus again decreases when $\alpha$ becomes $1$. When $\alpha$ becomes too large, the SAs put more weight only on the unlicensed band and the licensed bandwidth is wasted which reduces the user's surplus.  Surprisingly, our result shows that if $\alpha=0$, i.e., when the entire bandwidth $W$ is split between the SAs, the user's surplus is the lowest.  This suggests that if a regulator wants to increase the user surplus, it should enforce an $\alpha$ which is not too high nor too low and should select a moderate value of $\eta$. 
%Intuitively, when $L$ is small, in the equilibrium, both the SAs select the same ESC. Thus, the congestion price increases because of the same information. Hence, the user's surplus decreases. On the other hand, when $L$ is higher, a monopoly may arise where only SA $1$ has a positive profit. Thus, the user surplus again decreases when $L$ is high. Note that as $\alpha$ increases, the user's surplus increases, however, when $\alpha$ exceeds a threshold, the user's surplus decreases. Apparently, the congestion cost of SA $1$ is not minimized either at $\alpha=0$ or $\alpha=1$. Thus, the user's surplus increases initially, however, when $\alpha$ exceeds a threshold, the user's surplus decreases. Hence, randomization between unlicensed spectrum, and licensed spectrum increases the user's surplus.

\begin{figure}
%\end{minipage}\hfill
%\begin{minipage}{0.24\textwidth}
\includegraphics[trim=0in 0in 0in 0in,width=0.4\textwidth]{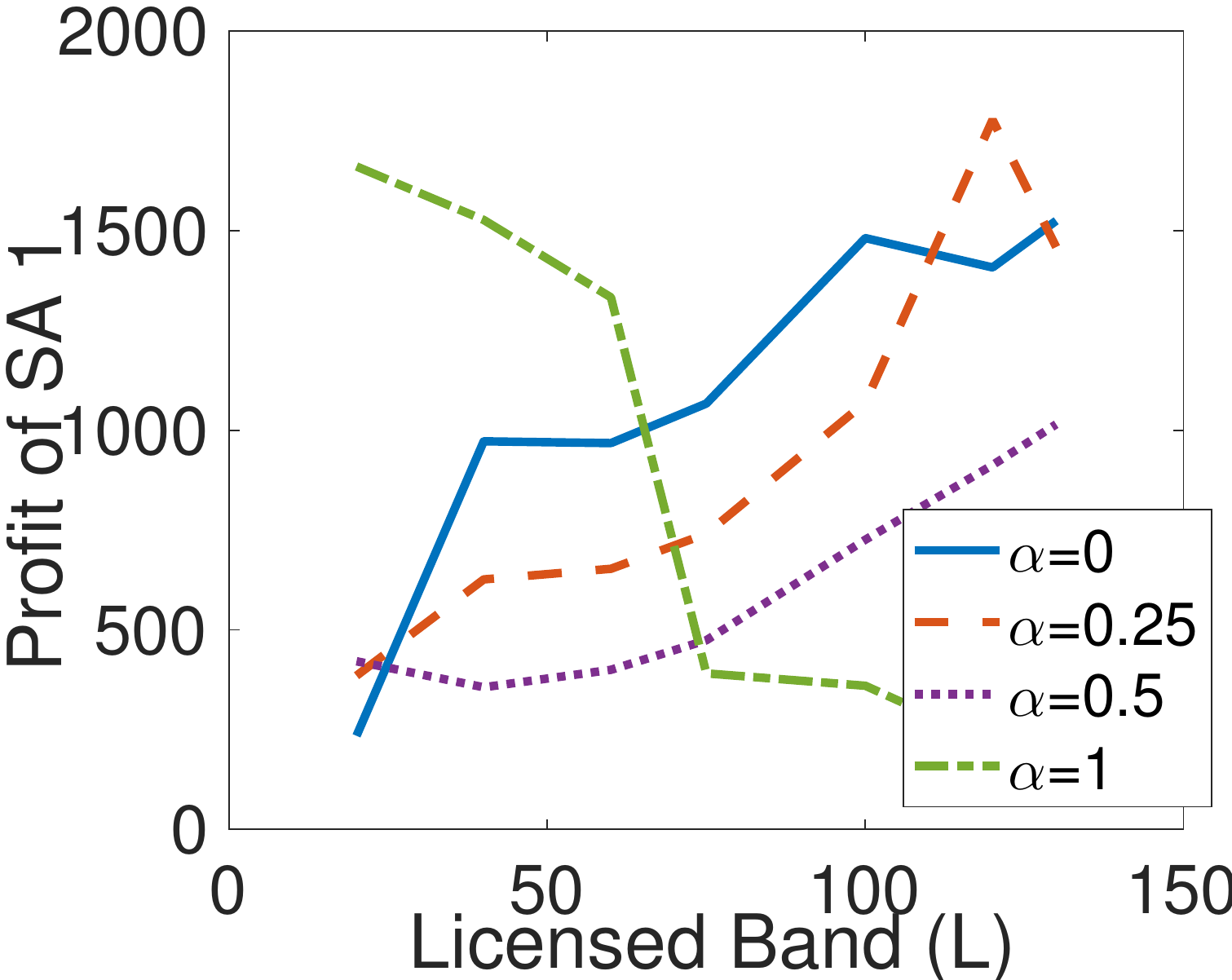}
\vspace{-.1in}
\caption{The profit of SA $1$ versus $L$ for different values of $\alpha$.}
\label{fig:pr1}
\vspace{-0.2in}
\end{figure}
\begin{figure}
%\end{minipage}\hfill
%\begin{minipage}{0.24\textwidth}
\includegraphics[trim=0in 0in 0in 0in,width=0.4\textwidth]{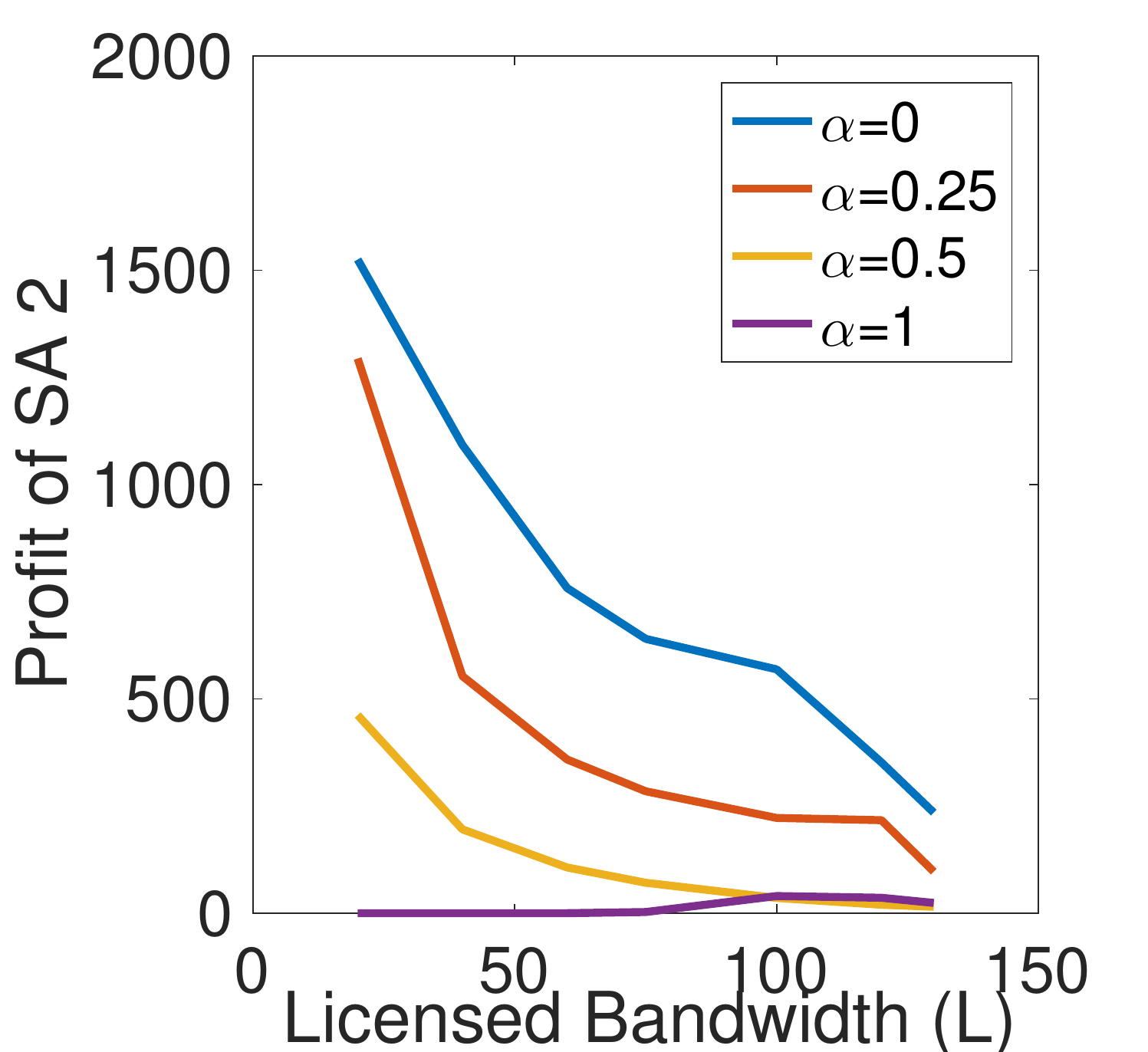}
\vspace{-0.1in}
\caption{The profit of SA $2$ versus  $L$ for different values of $\alpha$.}
\label{fig:pr2}
\vspace{-0.2in}
%\end{minipage}\hfill
\end{figure}

\subsection{Profits of SAs}
Figure~\ref{fig:pr1} shows that as $L$ increases, SA $1$'s payoff increases except when $\alpha=1$. Note that when $L$ is small, SA $1$'s congestion cost is higher, and so its profit  is small when $\alpha$ is also small. The congestion cost decreases as $L$ increases, hence, the SA $1$'s profit  also increases when $\alpha$ is small. However, as $L$ exceeds a threshold, the SA $1$'s profit decreases when $\alpha$ is small. Intuitively, when $\alpha$ is small, the SAs obtain information from the same ESC. SA $2$, thus, selects a very low price in order to have a positive subscription base when $L$ is large. Note from Figure~\ref{fig:pr2} that the payoff of SA $2$ decreases drastically once $L$ exceeds a threshold. However, if $L$ becomes too high and becomes almost equal to the entire bandwidth $W$, the profit of SA $1$ also increases since SA $1$ becomes a monopolist. 

 When $\alpha$ is high, SA $2$ obtains information from ESC $B$ when $L$ is high. Thus, SA $2$ always exists in the market. Thus, the profit of SA $1$ does not increase much even when $L$ is high. Note that when $\alpha=1$, SA $1$ only uses the unlicensed spectrum. Thus, the profit of SA $1$ decreases as $L$ increases since the unlicensed bandwidth decreases. Figure~\ref{fig:pr1} also shows that SA $1$ may not always prefer $\alpha=0$. SA $1$'s payoff is not always maximized at $\alpha=0$.
%This is because when $L$ is high, the number of customers served become small the profit of SA $1$ is higher when $\alpha$ is higher. Intuitively, when $L$ is small, the congestion cost in the licensed spectrum is the high. If $\alpha=0$, the SA $1$ has to use only the licensed spectrum, thus, the price of SA $1$ has to be low in order to attract customers. When $\alpha>0$, the SA can use the unlicensed spectrum more which increases the profit when $L$ is small. The profit becomes the highest when $\alpha=1$ when $L$ is small. However, as $L$ increases, the profit increases rapidly when $\alpha$ is small. This is because when $L$ is large, the congestion cost in the licensed band is low, thus, the SA $1$ can earn a larger profit by selecting $L$ with a higher probability. Note that when $\alpha=1$, the SA $1$ only uses the unlicensed spectrum, thus, the profit decreases as $L$ increases ($W-L$ decreases). 

%When $L$ is large, and $1>\alpha>0$, SAs select different ESCs in the equilibrium. Specifically, $\alpha\geq 0.25$ and $L>120$, in the equilibrium SA $1$ selects ESC $A$, and SA $2$ selects ESC $B$. However, when $L>120$ MHz, a monopoly scenario arises when $\alpha=0$. Thus, the profit of SA $1$ is higher when $\alpha=0$. 

Figure~\ref{fig:pr2} shows that as $\alpha$ increases, the profit of SA $2$ decreases. Intuitively, when $\alpha$ increases, the subscribers of SA $2$ face more congestion from SA $1$ as SA $2$ only uses the unlicensed spectrum. Thus, SA $2$ has to select a lower price which decreases the profit. The profit of SA $2$ also decreases as $L$ increases, as the unlicensed bandwidth is more congested for all values of $\alpha$. Note that when $\alpha=1$, the SAs become identical in nature. However, the only equilibrium is both the SAs obtain information from different ESCs. We assume that SA $2$ obtains information from ESC $B$ (i.e., it has an information of inferior quality). Thus, SA $2$'s payoff is strictly lower than that of SA $1$. 

%When $\alpha=1$ in an equilibrium SA $2$ obtains information from ESC $B$, and SA $1$ obtains information from ESC $A$. In this equilibrium, we have already shown that if $v$ exceeds a threshold the profit of SA $2$ is zero. The threshold decreases as $W-L$ increases. Thus, if $W-L$ is high, the profit of SA $2$ is zero. 

\begin{figure}
%\begin{minipage}{0.24\textwidth}
\includegraphics[trim=0in 0in 0in 0in,width=0.4\textwidth]{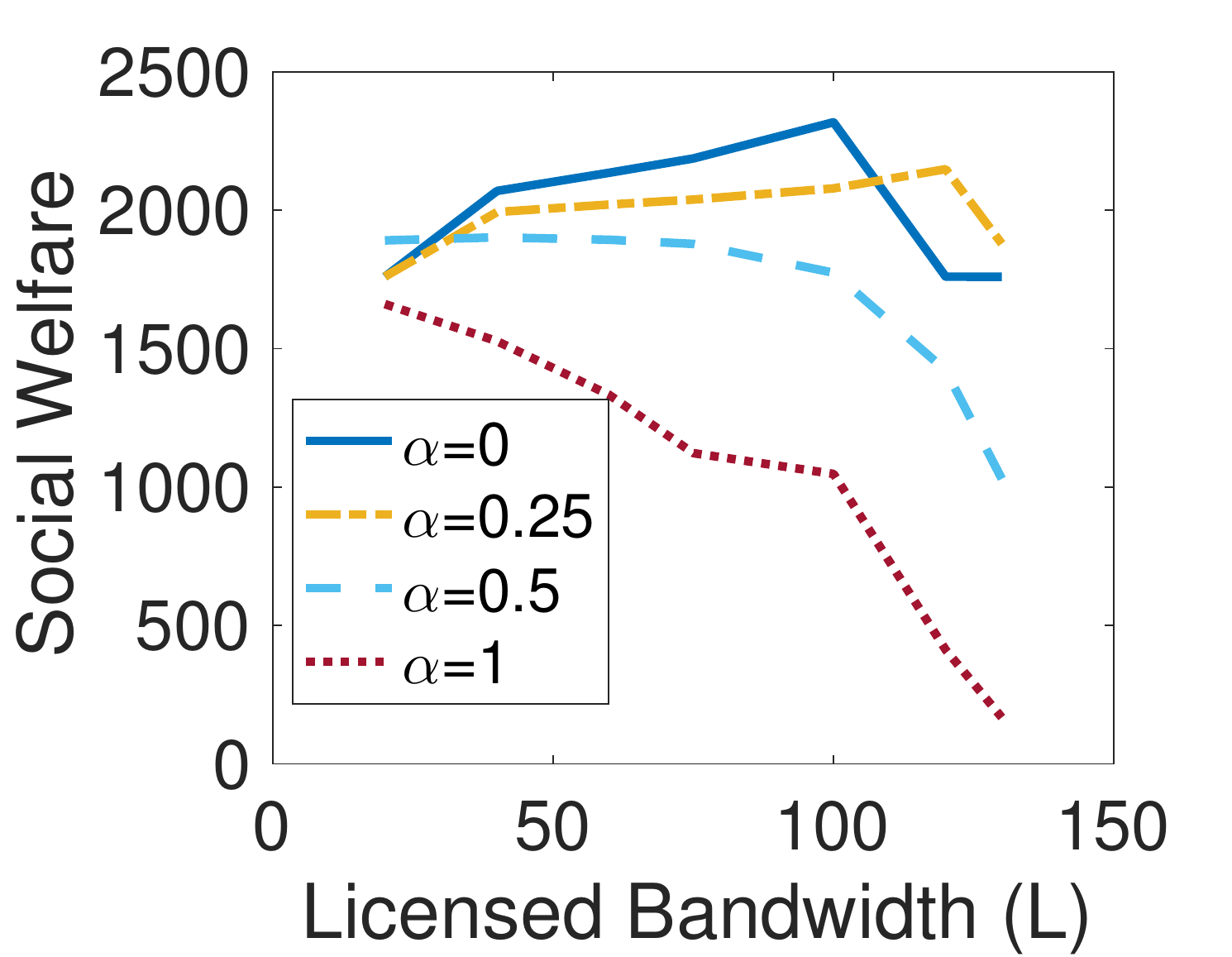}
\caption{Social Welfare versus $L$ for different values of $\alpha$.}
\label{fig:soc_wel}
%\end{minipage}
\vspace{-0.2in}
\end{figure}

\subsection{Social Welfare}
We denote the social welfare as the sum of the user surplus and the profits of the SAs. 
Figure~\ref{fig:soc_wel} shows that the social welfare increases initially as $L$ increases when $\alpha<0.5$.   The social welfare decreases when $L$ exceeds a threshold.  This is because the profit of SA $1$ and user surplus both increase initially as $L$ increases when $\alpha<0.5$. However, when $L$ exceeds a threshold, the user's surplus and the profits of SAs decrease. Thus, a higher value of $L$ decreases the social welfare.  Note that the increase of social welfare with $L$ is slow as $\alpha$ increases. This is because when $L$ increases, the increase of profit of SA $1$ is slower when $\alpha$ is higher. 

 When $\alpha=0.5$, social welfare almost remains constant when $L$ is below a threshold. When $L$ exceeds a threshold, the social welfare decreases. The increase of SA $1$'s profit with $L$ is smaller as $\alpha$ increases. Thus, the decrease of SA $2$'s profit with $\alpha$ nullifies the increase in the profit of SA $1$, and the increase of the user's surplus. When $L$ exceeds a threshold, the user's surplus also decreases. Hence, the social welfare decreases when $L$ exceeds a threshold. 
 When $\alpha=1$, the social welfare always decreases as $L$ increases. This is because when $\alpha=1$, the SAs only use the unlicensed spectrum. Thus, the profit of SA $1$ decreases drastically as $L$ increases. 
  %Our analysis shows that $L$ can not be too large or too small to maximize the social welfare. Note from Fig.~\ref{fig:pr2} that the profit of SA $2$ increases when $L$ is higher. However, the decreases in the SA $1$ nullifies the effect of increment in profit of SA $2$. 
  
  Figure~\ref{fig:soc_wel} shows that social welfare is not maximized at very high values of $L$ or at low values of $L$. Social welfare may not be maximized at extreme values of $\alpha$ either. When the ratio $\eta$ is moderate, the social welfare is the highest. However, the value of $\alpha$ that maximizes the social welfare will depend on the ratio $\eta$. %Smaller values of $\alpha$ is beneficial for social welfare, however, the exact value of $\alpha$ will depend on the $L$. 

%% file: proof.tex
\appendix
\section{Proof of Theorem 1}
We, first, characterize the price strategy when $\lambda_1<\Lambda$, subsequently, we characterize the price strategy when $\lambda_1=\Lambda$. 

If $\lambda_1<\Lambda$,  then, 
\begin{align}
q_jv-q_j\dfrac{\alpha^2\lambda_1}{W-L}-q_j\dfrac{(1-\alpha)^2\lambda_1}{L}-p_1=0
\end{align}
Now, using the above expression,
\begin{align}
p_1\lambda_1=p_1\dfrac{q_jv-p_1}{q_j\dfrac{\alpha^2}{W-L}+q_j\dfrac{(1-\alpha)^2}{L}}
\end{align}
Thus, the optimal $p_1^{*}=q_jv/2$. Hence, 
\begin{align}
\lambda_1=\dfrac{v/2}{\dfrac{\alpha^2}{W-L}+\dfrac{(1-\alpha)^2}{L}}
\end{align}
Note that $p_1=q_jv-q_j\dfrac{\alpha^2\lambda_1}{W-L}-q_j\dfrac{(1-\alpha)^2\lambda_1}{L}$. 

$\lambda_1$ can be at most $\Lambda$. Thus, if $\lambda_1=\Lambda$, the  price is
\begin{align}
p_1=q_jv-q_j\dfrac{\alpha^2\Lambda}{W-L}-q_j\dfrac{(1-\alpha)^2\Lambda}{L}
\end{align}
Note that this is the highest possible price with $\lambda_1=\Lambda$. Hence, the result follows.

\section{Proof of Theorem 2}
%From Theorem 3 and Lemma 2 which have been proved independent of Theorem 2, note that if Suppose $p_2=0$. 

First, suppose that $\lambda_2=0$ and $p_2=0$. Note that if $\lambda_1<\Lambda$, the user's surplus must be zero. If the user's surplus is same as in SA $2$, then
\begin{align}\label{eq:lamb1}
\lambda_1=\min\{\dfrac{v(W-L)}{\alpha},\Lambda\}
\end{align}
and 
\begin{align}\label{pr_eq}
p_1=q_jv-q_j\alpha^2\dfrac{\lambda_1}{W-L}-q_j(1-\alpha)^2\dfrac{\lambda_1}{L}
\end{align}
Now, we show that it is an NE. 

If the user's surplus from SA $2$ as same as in SA $1$, then we have 
\begin{align}\label{in2}
(1-\alpha)\dfrac{\alpha\lambda_1+\lambda_2}{W-L}-\dfrac{(1-\alpha)^2\lambda_1}{L}=p_1-p_2
\end{align}
Since $\dfrac{W-L}{L}\leq \dfrac{2\alpha-1}{2(1-\alpha)}$ and $\dfrac{\alpha}{2(1-\alpha)}\geq \dfrac{2\alpha-1}{2(1-\alpha)}$, thus, $\dfrac{W-L}{L}\leq \dfrac{\alpha}{2(1-\alpha)}$. Thus, we obtain
\begin{align}
(1-\alpha)\dfrac{\alpha\lambda_1+\lambda_2}{W-L}-\dfrac{(1-\alpha)^2\lambda_1}{L}\nonumber\\
\leq \dfrac{2(1-\alpha)^2\lambda_1}{L}-\dfrac{(1-\alpha)^2\lambda_1}{L}\nonumber\\
=\dfrac{(1-\alpha)^2\lambda_1}{L}
\end{align}
the inequality follows as $\lambda_2=0$. Hence, from (\ref{in2}) $\lambda_1$ increases as $p_1$ increases. The highest possible $p_1$ is given by (\ref{pr_eq}). 

If $\lambda_1=\Lambda$, then $\lambda_1$ is replaced b $\Lambda$ in the expression of $p_1$. 

Now, we are only left to show that $\lambda_2=0$ and $p_2=0$ when $\dfrac{W-L}{L}\geq \dfrac{2\alpha-1}{2(1-\alpha)}$. We obtain the above in Theorem 3 and Lemma 2 which we have proved independent of the above theorem. Finally, note that the result where Hence, the result follows. 
%If the price $p_1$ is such that
%\begin{align}
%p_1=q_jv-q_j\alpha^2\dfrac{\lambda_1}{W-L}-q_j(1-\alpha)^2\dfrac{\lambda_1}{L}
%\end{align}
%where $\lambda_1$ is given by (\ref{eq:lamb1}) then $\lambda_2=0$. If $p_1$ is other than any value it will not increase the payoff. This is because if $\lambda_2=0$, the optimal price is the monopoly price $q_jv/2$ if $\lambda_1<\Lambda$. However, since
%\begin{align}
%\dfrac{W-L}{L}\leq \dfrac{2\alpha-1}{2(1-\alpha)}
%\end{align}
%We, first, show that the best response of SA $1$ is as stated in Theorem 2. First, consider the
%\begin{align}
%(1-\alpha)\dfrac{\alpha\lambda_1+\lambda_2}{W-L}-\dfrac{(1-\alpha)^2\lambda_1}{L}=p_1-p_2
%\end{align}

\section{Proof of Theorem 3}
First consider the scenario that $\lambda_1+\lambda_2=\Lambda$, and both $\lambda_1>0,
\lambda_2>0$. 

Since both $\lambda_1>0, \lambda_2>0$, the user's surplus from both the SAs is positive and equal, thus,  
\begin{align}
(1-\alpha)\dfrac{\alpha\lambda_1+\lambda_2}{W-L}-\dfrac{(1-\alpha)^2\lambda_1}{L}=p_1-p_2
\end{align}
Replacing $\lambda_2=\Lambda-\lambda_1$, we have
\begin{align}\label{eq:lambda1}
& (1-\alpha)\dfrac{\alpha\lambda_1+\Lambda-\lambda_1}{W-L}-\dfrac{(1-\alpha)^2\lambda_1}{L}=p_1-p_2\nonumber\\
& \dfrac{(1-\alpha)^2\lambda_1}{W-L}+\dfrac{(1-\alpha)^2\lambda_1}{L}=\dfrac{(1-\alpha)\Lambda}{W-L}+p_2-p_1
\end{align}
SA $1$'s payoff is
\begin{align}
p_1\lambda_1
\end{align}
From (\ref{eq:lambda1}), we obtain the optimal $p_1^{*}$ from first principle as,
\begin{align}\label{eq:p1}
p_1^{*}=\dfrac{(1-\alpha)\Lambda}{2(W-L)}+p_2/2.
\end{align}
it is also unique since $p_1\lambda_1$ is strictly concave in $p_1$.

Similarly in (\ref{eq:lambda1}) replacing $\lambda_1$ with $\Lambda-\lambda_2$, we obtain 
\begin{align}\label{eq:lambda2}
& \dfrac{(1-\alpha)^2\lambda_2}{W-L}+\dfrac{(1-\alpha)^2\lambda_2}{L}=\dfrac{(1-\alpha)^2\Lambda}{L}-\dfrac{(1-\alpha)\alpha\Lambda}{W-L}+p_1-p_2
\end{align}
Hence, optimal $p_2^{*}$ is given by
\begin{align}\label{eq:p2}
p_2^{*}=\dfrac{(1-\alpha)^2\Lambda}{2L}-\dfrac{(1-\alpha)\alpha\Lambda}{2(W-L)}+p_1^{*}/2
\end{align}
Hence, from (\ref{eq:p1}) and (\ref{eq:p2}), we obtain $p_1^{*}$ and $p_2^{*}$. $\lambda_1$ and $\lambda_2$ are obtained from (\ref{eq:lambda1}) and (\ref{eq:lambda2}) once $p_1^{*}$ and $p_2^{*}$ are found. 

Note that, the user's surplus has to be non-negative which we did not consider so-far. Since both $\lambda_i>0$, thus, the user's surplus is the same from both the SAs. Hence, we must have
\begin{align}
q_jv\geq q_j\dfrac{\alpha\lambda_1+\lambda_2}{W-L}+p_2^{*}
\end{align}
The right hand side of the expression gives the value of $\beta(\alpha)$.

Note that $p_2^{*}<0$ if $\dfrac{W-L}{L}<\dfrac{2\alpha-1}{2(1-\alpha)}$, thus, the equilibrium is not sustainable when $\dfrac{W-L}{L}<\dfrac{2\alpha-1}{2(1-\alpha)}$. 

%%% Theorem 4 %%%%
\section{Proof of Theorem 4}
In order to prove Theorem 4, we, first, show the following lemma
\begin{lem}
If $\frac{W-L}{L}>\frac{\alpha}{2(1-\alpha)}$, and$v<\beta(\alpha)$, the equilibrium is the following
\begin{align}
p_1& =\dfrac{q_jv(1-\alpha)[\dfrac{(1-\alpha)(2-\alpha)}{L}+\dfrac{\alpha^2}{W-L}]}{\dfrac{4(1-\alpha)^2}{L}+\dfrac{3\alpha^2}{W-L}}\nonumber\\
p_2& =\dfrac{q_jv(1-\alpha)[\dfrac{2(1-\alpha)}{L}-\dfrac{\alpha}{W-L}]}{\dfrac{4(1-\alpha)^2}{L}+\dfrac{3\alpha^2}{W-L}}.
\end{align}
The third stage wardrop equilibrium is
\begin{align}\label{eq:lambda}
\lambda_1& =\dfrac{p_1}{q_j(1-\alpha)^2/L},\nonumber\\
\lambda_2& =\dfrac{p_2(\alpha^2/(W-L)+(1-\alpha)^2/L)}{q_j(1-\alpha)^2/L(W-L)}.
\end{align}
\end{lem}
{\em proof}: We have characterized the equilibrium when $\lambda_1+\lambda_2=\Lambda$ in Theorem 3 and it is an equilibrium only when $v>\beta(\alpha)$. We will now characterize the equilibrium when $\lambda_1+\lambda_2<\Lambda$. 

Note that when $\lambda_1+\lambda_2<\Lambda$, the user's surplus is zero. Hence, we have the following
\begin{align}\label{eq:pr1}
q_jv-q_j\dfrac{\alpha^2\lambda_1+\alpha\lambda_2}{W-L}-q_j\dfrac{(1-\alpha)^2\lambda_1}{L}=p_1.
\end{align}
If $\lambda_2>0, \lambda_1>0$, the user's surplus from both the SAs is equal. Hence, 
\begin{align}\label{eq:diff1}
(1-\alpha)\dfrac{\alpha\lambda_1+\lambda_2}{W-L}-\dfrac{(1-\alpha)^2\lambda_1}{L}=p_1-p_2
\end{align}
From (\ref{eq:pr1}) and (\ref{eq:diff1}), one can solve for $\lambda_1$ and $\lambda_2$ in terms of $p_1$ and $p_2$. Thus, one can obtain optimal $p_1^{*}$ in terms of $p_2^{*}$. Hence, we obtain the equilibrium price strategy $p_1^{*}$ and $p_2^{*}$. The rest of the proof is algabraic, and hence, we remove it. \qed

Now, we show Theorem 4. Note that the price of SA $2$ is positive as long as $\dfrac{2(1-\alpha)}{L}-\dfrac{\alpha}{W-L}>0$ and $v<\beta(\alpha)$. We have already shown that if $v\geq \beta(\alpha)$, the prices of SAs are positive if $\dfrac{W-L}{L}\geq \dfrac{2\alpha-1}{2(1-\alpha)}$. Since $\dfrac{\alpha}{2(1-\alpha)}>\dfrac{2\alpha-1}{2(1-\alpha)}$. Thus, $\dfrac{W-L}{L}\geq \dfrac{\alpha}{2(1-\alpha)}\geq \dfrac{2\alpha-1}{2(1-\alpha)}$. Hence, the equilibrium prices are positive. Hence, the result follows.

%%% Theorem 5%%%
\section{Proof of Theorem 5}
Before proving Theorem 5, we, first, introduce the following two lemmas
\begin{lem}\label{lem3}
The only equilibrium where $\lambda_1+\lambda_2=\Lambda$ is the following, 
\begin{align}\label{eq:1a2bpr}
p_1=\dfrac{(q_A-q_B)v}{3}+\dfrac{(q_A\alpha^2-3q_B\alpha+2q_B)\Lambda}{3(W-L)}+\dfrac{q_A(1-\alpha)^2\Lambda}{3L}\nonumber\\
p_2=\dfrac{(q_B-q_A)v}{3}+\dfrac{(2q_A\alpha^2-3q_B\alpha+q_B)\Lambda}{3(W-L))}+\dfrac{2q_A(1-\alpha)^2\Lambda}{3L}
\end{align}
Under the second stage pricing strategy profile, the third stage Wardrop equilibrium is
\begin{align}
\lambda_1=\dfrac{p_1}{\dfrac{q_A\alpha^2}{W-L}+\dfrac{q_A(1-\alpha)^2}{L}+\dfrac{q_B(1-2\alpha)}{W-L}}\nonumber\\
\lambda_2=\dfrac{p_2}{\dfrac{q_A\alpha^2}{W-L}+\dfrac{q_A(1-\alpha)^2}{L}+\dfrac{q_B(1-2\alpha)}{W-L}}
\end{align}
and 
\begin{align}\label{eq:con}
\dfrac{(q_A-q_B)v}{3}\leq \dfrac{(2q_A\alpha^2-3q_B\alpha+q_B)\Lambda}{3(W-L))}+\dfrac{2q_A(1-\alpha)^2\Lambda}{3L}
\end{align}
\begin{align}\label{eq:condn1a2b}
q_Bv\geq p_2+q_B\alpha\lambda_1/(W-L)+q_B\lambda_2/(W-L)
\end{align}
%\end{align}
\end{lem}
{\em Proof}: Since $\lambda_1>0, \lambda_2>0$, the user's surplus must be the same from the SAs. Thus, we have
\begin{align}
& (q_A-q_B)v-(q_A-q_B)\dfrac{\alpha^2\lambda_1}{W-L}+q_B(1-\alpha)\dfrac{\alpha\lambda_1+\lambda_2}{W-L}\nonumber\\ & -q_A\dfrac{(1-\alpha)^2\lambda_1}{L}=p_1-p_2
\end{align}
Since $\lambda_1+\lambda_2=\Lambda$, we obtain both $\lambda_1, \lambda_2$ in terms of $p_1,p_2$ similar to the one we obtained while proving Theorem 3. Hence, we obtain optimal $p_1^{*},p_2^{*}$ since the expression $p_i\lambda_i$ is concave in $p_i$. The rest of the proof is algabraic, and we omit it here. 

Note that $p_2$ is only positive when (\ref{eq:con}) is satisfied. On the other hand, we must have the user's surplus is positive which leads to the inequality in (\ref{eq:condn1a2b}). Hence, the result follows. \qed

We, now, state another Lemma which we use to prove Theorem 5. 
\begin{lem}\label{lem4}
The only equilibrium where $\lambda_1+\lambda_2<\Lambda$, if 
\begin{align}\label{ineq:pos1}
 \frac{W-L}{L}\geq \frac{q_B\alpha^2/q_A+\alpha-2\alpha^2}{2(1-\alpha)^2},
\end{align}
and the second stage pricing strategy is 
\begin{align}
p_1=  v\dfrac{\dfrac{2q_A^2\alpha^2-q_Aq_B\alpha^3-q_Aq_B\alpha^2}{W-L}+\dfrac{(2q_A^2-q_Bq_A\alpha)(1-\alpha)^2}{L}}{4q_A[\alpha^2/(W-L)+(1-\alpha)^2/L]-q_B\alpha^2/(W-L)]}\nonumber\\
p_2=v\dfrac{\dfrac{2q_Aq_B\alpha^2-q_Bq_A\alpha-q_B^2\alpha^2}{W-L}+\dfrac{2q_Aq_B(1-\alpha)^2}{L}}{4q_A[\alpha^2/(W-L)+(1-\alpha)^2/L]-q_B\alpha^2/(W-L)]}
\end{align}
and the condition in (\ref{eq:condn1a2b}) is not satisfied. 
\end{lem}
{\em Proof}: We proceed similar to Lemma 2. When $\lambda_1>0, \lambda_2>0$, the user's surplus must be equal for both the SAs. Hence, 
\begin{align}
& (q_A-q_B)v-(q_A-q_B)\dfrac{\alpha^2\lambda_1}{W-L}+q_B(1-\alpha)\dfrac{\alpha\lambda_1+\lambda_2}{W-L}-\nonumber\\ & q_A\dfrac{(1-\alpha)^2\lambda_1}{L}=p_1-p_2
\end{align}
Since $\lambda_1+\lambda_2<\Lambda$, thus, we must have the user's surplus as zero. Hence,
\begin{align}
q_Av-(q_A-q_B)\dfrac{\alpha^2\lambda_1}{W-L}-q_B\dfrac{\alpha^2\lambda_1+\alpha\lambda_2}{W-L}-q_A\dfrac{(1-\alpha)^2\lambda_1}{L}=p_1
\end{align}
We obtain $\lambda_1,\lambda_2$ in terms of $p_1, p_2$ from the above two expressions. We, thus, obtain the optimal $p_i^{*}$ $i=1,2$ by differentiating  $p_i\lambda_i$ since $p_i\lambda_i$ is strictly concave in $p_i$. Note that optimal $p_i^{*}$ is a function of both $p_1,p_2$. Thus, by replacing $p_2^{*}$ in $p_1^{*}$ we obtain the equilibrium price $p_1^{*}$ devoid of $p_2^{*}$. Similarly, we obtain the equilibrium price $p_2^{*}$. Since the steps are algabraic, we omit it here. 

Note that the price of $p_2$ is only positive when the inequality in (\ref{ineq:pos1}) is satisfied. When the inequality in (\ref{eq:condn1a2b}) is not satisfied $\lambda_1+\lambda_2<\Lambda$ as otherwise we will operate in the regime of Lemma 3. Hence, result follows. \qed

Now, we prove Theorem 5. Note from Lemma~\ref{lem3} that the price strategy of SA $2$ is positive as long as the condition in (\ref{eq:condn1ahighv}) is satisfied. Note from Lemma~\ref{lem4} that the price strategy of SA $2$ is positive as long as the condition in (\ref{ineq:pos}) is satisfied. The price strategy of SA $1$ is always positive from Lemmas~\ref{lem4} and \ref{lem3}. Hence, the result follows. \qed
%Note from the price strategy if the inequality in (\ref{ineq:pos}) is satisfied, $p_2$ is positive. Hence, the result follows. 

%%% Theorem 6%%%%%
\section{Proof of Theorem 6}
We again start with two Lemmas which we use to prove Theorem 6. 
\begin{lem}\label{thm:highlambda}
Consider the following pricing strategy 
\begin{align}\label{eq:pr1b2a}
p_1&=\frac{(q_B-q_A)v}{3}+\dfrac{(2q_A-3q_B\alpha+q_B\alpha^2)\Lambda}{3(W-L)}+\frac{q_B(1-\alpha)^2\Lambda}{3L}\nonumber\\
p_2 &=\frac{(q_A-q_B)v}{3}+\dfrac{(2q_B\alpha^2-3q_B\alpha+q_A)\Lambda}{3(W-L)}+\frac{2q_B(1-\alpha)^2\Lambda}{3L}
\end{align}
Under the above pricing strategy, the third stage Wardrop equilibrium is
\begin{align}
\lambda_1=\dfrac{p_1}{\dfrac{q_B\alpha^2}{W-L}+\dfrac{q_B^2(1-\alpha)^2}{L}\dfrac{q_A-2q_B\alpha}{W-L}}\nonumber\\
\lambda_2=\dfrac{p_2}{\dfrac{q_B\alpha^2}{W-L}+\dfrac{q_B^2(1-\alpha)^2}{L}\dfrac{q_A-2q_B\alpha}{W-L}}\nonumber
\end{align}
The pricing strategy stated in (\ref{eq:pr1b2a}) is the unique Nash equilibrium if the following hold
\begin{align}
(q_A-q_B)v\leq\dfrac{(2q_A-3q_B\alpha+q_B\alpha^2)\Lambda}{3(W-L)}+\dfrac{q_B(1-\alpha)^2\Lambda}{3L};
\end{align}
and 
%\begin{align}
\begin{align}\label{eq:ineq2}
q_Av\geq p_2+\dfrac{q_B\alpha\lambda_1}{W-L}+q_A\dfrac{\lambda_2}{W-L}.
\end{align}
%\end{align}
and the above price strategy is the only one where $\lambda_1+\lambda_2=\Lambda$. 
\end{lem}
Thus, if the inequality in (\ref{eq:pricehighv1b}) the price of SA $1$ must be $0$. The proof of the above lemma is similar to Lemma~\ref{lem3}. Hence, we omit it here.\qed

\begin{lem}\label{lem6}
The unique second stage pricing strategy is
\begin{align}
p_1=\dfrac{v(\dfrac{2q_Bq_A\alpha^2-q_Bq_A\alpha^3-q_B^2\alpha^2}{W-L}+\dfrac{(1-\alpha)^2(2-\alpha)q_Aq_B}{L})}{4q_A[\alpha^2/(W-L)+(1-\alpha)^2/L]-q_B\alpha^2/(W-L)}\nonumber\\
p_2=\dfrac{v(\dfrac{2q_A^2\alpha^2-q_Aq_B\alpha^2-q_Aq_B\alpha}{W-L}+\dfrac{2q_A^2(1-\alpha)^2}{L})}{4q_A[\alpha^2/(W-L)+(1-\alpha)^2/L]-q_B\alpha^2/(W-L)}
\end{align}
if 
\begin{align}\label{ineq:posi}
\dfrac{W-L}{L}\geq \dfrac{q_B\alpha^2+q_B\alpha-2q_A\alpha^2}{2q_A(1-\alpha)^2}
\end{align}
and $\lambda_1+\lambda_2<\Lambda$.
\end{lem}
{\em proof}: The proof of the above lemma is similar to the proof of Lemma~\ref{lem4}. Hence, we omit it here.\qed

Now, we show Theorem 6. Note from Lemma~\ref{thm:highlambda} that $p_1$ is positive in the inequality in (\ref{eq:pricehighv1b}) is satisfied. Note that $p_2>0$ only when the inequality in (\ref{ineq:pos1}) is satisfied by Lemma~\ref{lem6}. When $\lambda_1+\lambda_2<\Lambda$, the price strategy $p_1$ is always positive. Also note that when $\lambda_1+\lambda_2=\Lambda$, the price strategy $p_2$ is always positive from Lemma~\ref{thm:highlambda}. Hence,  the result follows. \qed
%First consider the scenario that $\lambda_1+\lambda_2=\Lambda$, and both $\lambda_1>0,
%\lambda_2>0$. Thus, 

%%% Theorem7%%%%
\section{Proof of Theorem 7}
Note that if $\alpha=1$, and both the SAs obtain information from the same ESC, the prices of both the SAs become zero. However, SAs have pay the ESCs.  Hence, there is no equilibrium where both the SAs obtain information from the same ESC when $\alpha=1$. 

On the other hand if $\alpha=0$, and suppose that SA $i$ obtains information from ESC $B$, and SA $j$, $j\neq i$ obtains information from ESC $A$. Then, SA $i$ can obtain strictly higher payoff by obtaining information from ESC $A$. Hence, there is no equilibrium where the SAs obtain information from different ESCs when $\alpha=0$. 

%%% Theorem 8%%%
\section{Proof of Theorem 8}
Note that when $\alpha<1$,and$\dfrac{W-L}{L}\rightarrow \infty$, the equilibrium prices of both the SAs is positive by Theorem 4. Further, when $\beta(\alpha)\rightarrow \infty$ when $\dfrac{W-L}{L}\rightarrow \infty$. Hence, by Lemma 2, the payoff of each of the SAs when they obtain information from the same ESC $A$ is given by if $\alpha<1$,
\begin{align}
\pi_1(A,A)=Lq_A\dfrac{v^2(2-\alpha)^2}{16(1-\alpha)^2}-\tilde{p}_A\nonumber\\
\pi_2(A,A)=(W-L)q_A\dfrac{v^2(1-\alpha)^2}{4}-\tilde{p}_A
\end{align}

On the other hand, if SA $1$ obtains information from ESC $A$, and SA $2$ obtains information from ESC $B$, the payoffs are
\begin{align}
\pi_1(A,B)=L\dfrac{v^2(2q_A-q_B\alpha)^2}{16q_Aq_B(1-\alpha)^2}-\tilde{p}_A\nonumber
\pi_2(A,B)=(W-L)q_B\dfrac{v^2(1-\alpha)^2}{4}-\tilde{p}_B
\end{align}
Since $q_B<q_A$, and $\tilde{p}_A$  is small enough, thus, the SA $2$ will not obtain information from ESC $B$, rather, it can attain a higher payoff by obtaining information from ESC $A$.

If SA $1$ obtains information from ESC $B$, and SA $2$ obtains information from ESC $A$, the payoffs are
\begin{align}
\pi_1(B,A)=Lq_B\dfrac{v^2(2-\alpha)^2}{16(1-\alpha)^2}-\tilde{p}_B\nonumber\\
\pi_2(B,A)=(W-L)q_A\dfrac{v^2(1-\alpha)^2}{4}-\tilde{p}_A
\end{align}
Since $q_B<q_A$, and $\tilde{p}_A$ is small enough,  the SA $1$ will have a strictly higher payoff by obtaining information from ESC $A$ rather from ESC $B$ when SA $2$ obtains information from ESC $A$. Hence, the strategy profile where both the SAs obtain information ESC $A$ is a Nash equilibrium when $\tilde{p}_A$ is small enough. Note that it is the unique Nash equilibrium since the strategy profile where both the SAs obtain information from ESC $B$ is not sustainable when $\tilde{p}_A$ is small. \qed

Also note that if $\tilde{p}_A$ or $\tilde{p}_B$ is not small, SA $2$ may have a monopoly in this scenario if $L$ if $\alpha$ is small enough or $L$ is very small.

%The rest of the proof follows from the fact that as $\dfrac{W-L}{L}\rightarrow \infty$, the payoff of SAs will be higher if they obtain information from the ESC $A$ rather than obtaining information from different ESCs when $\tilde{p}_A$ is small enough. 

%%% Theorem 9%%%
\section{Proof of Theorem 9}
If $\dfrac{W-L}{L}\rightarrow 0$, note from Theorems 3 and Theorem 4 that the price strategy of SA $2$ is zero when the SAs will obtain information from the same ESC if $\alpha>0$. Thus, the equilibrium where SAs will obtain information from the same ESC is not sustainable. 

Note from  Theorem 5 that if $0<\alpha<1/2$, the price strategy of SA $2$ is zero when SA $2$ obtains information from ESC $B$ and SA $1$ obtains information from ESC $A$.  Only when $\dfrac{W-L}{L}\geq \dfrac{q_B\alpha^2/q_A+\alpha-2\alpha^2}{2(1-\alpha)^2}$, SA $2$ selects a positive price in the equilibrium. From Corollary 1 (in Section 5.1), there is an $\alpha$ for which the price strategy of SA $2$ is positive. Hence, SA $2$ attains a positive profit if $\tilde{p}_B$ is small when SA $1$ obtains information from ESC $A$, and SA $2$ obtains information from ESC $B$. \qed

%  that if $0<\alpha<1/2$, and $\dfrac{W-L}{L}\rightarrow 0$, SA $2$'s price becomes zero irrespective from which ESC SA $2$ obtains information from  if SA $1$ obtains information from ESC $A$. On the other hand, from Theorem 6, the price of SA $1$ becomes zero if SA $1$ obtains information from ESC $B$, and SA $2$ obtains information from ESC $A$ if $0<\alpha<1/2$, and $\dfrac{W-L}{L}\rightarrow 0$. Thus, both the SAs will remain active is not possible if $0<\alpha<1/2$, and $\dfrac{W-L}{L}\rightarrow 0$. Now, SA $2$ can not have a monopoly power since SA $1$ can obtain information from ESC $A$ and obtains a positive payoff since $\tilde{p}_A$ is small enough. Thus, the only possibility is that SA $1$ will have a monopoly power. 
%
%If $\alpha>1/2$ and from Corollary 1 (in Section 5.1), SA $2$ obtains a strictly higher payoff by obtaining information from ESC $B$ compared to obtaining information from ESC $A$ when SA $1$ obtains information from ESC $A$. Thus, the strategy where SA $1$ obtains information from ESC $A$ and SA $2$ obtains information from ESC $B$ is sustainable when $\tilde{p}_A, \tilde{p}_B$ are small enough. Hence, the result follows. 